\documentclass[preprint ,12pt]{elsarticle}
\usepackage{amssymb}

\begin{document}

\begin{frontmatter}

\title{Extragalactic propagation of ultrahigh energy cosmic-rays}

\author{Denis Allard}

\address{Laboratoire Astroparticule et Cosmologie (APC), Universit\'e Paris 7/CNRS, 10 rue A. Domon et L. Duquet, 75205 Paris Cedex 13, France.}

\begin{abstract}
%
In this paper we review the extragalactic propagation of ultrahigh energy cosmic-rays (UHECR). We present the different energy loss processes of protons and nuclei, and their expected influence on  energy evolution of the UHECR spectrum and composition. We discuss the possible implications of the recent composition analyses provided by the Pierre Auger Observatory. The influence of extragalactic magnetic fields and possible departures from the rectilinear case are also mentioned as well as the production of secondary cosmogenic neutrinos and photons and the constraints their observation would imply for the UHECRs origin. Finally, we conclude by briefly discussing the relevance of a multi messenger approach for solving the mystery of UHECRs.
\end{abstract}


\end{frontmatter}


\section{Introduction}
\label{sec:intro}

After more than fifty years of experimental efforts, the origin of ultrahigh energy cosmic-rays (UHECRs, e.g, cosmic-rays above $\sim 10^{18}$ eV) remains a mystery. The understanding of the production of these particles, the most energetic particles in the universe, is one of the most intense research fields of high energy astrophysics. Since the pioneering experiment at Volcano Ranch  and the observation of the first cosmic-ray event above $10^{20}$ eV (see \cite{NaganoWatson} for a complete review of the early experiments),  large statistics have been accumulated above $10^{18}$ eV.  High resolution measurements of the UHECR spectrum, composition, and arrival direction have been allowed by recent experiments like AGASA \cite{Nagano92},  HiRes \cite{HiRes},  the Pierre Auger Observatory \cite{Augerer2004}, and Telescope Array \cite{TA}. Among the most interesting recent results, one can cite (see \cite{Sommers}) the evidence for a suppression of the UHECR flux above 3-5$\times10^{19}$ eV observed by HiRes \cite{HiResCut,Cut} and Auger \cite{AugerSpec2008} with a large significance. Furthermore, the recent analyses at the Pierre Auger Observatory seem to indicate an evolution of the composition toward heavier elements above the ankle \cite{Augerer2010} as well as hints for an anisotropic distribution of the arrival directions of the highest energy events \cite{AugerScience} and in particular a possible diffuse excess in the direction of the Centaurus constellation\cite{AugerUpdate}. Since the statistics above $\sim3\times10^{19}$ eV are quite low, and the consistency between the results of different experiments is still a matter of debate, these trends remain to be confirmed and understood with future data.

The extragalactic origin of UHECRs (at least above the ankle of the cosmic-ray spectrum) is widely accepted. As a consequence the measured cosmic-ray spectrum on Earth has to be shaped by the effect of the propagation of the particles in the extragalactic medium. During their journey from the source to the Earth the injected cosmic-ray spectrum and composition can be modified by interactions with photon backgrounds and cosmic magnetic fields. A detailed modeling of the extragalactic propagation of UHECRs is then a necessary ingredient for the astrophysical interpretation of the data. One of the most important features due to UHECRs extragalactic propagation is the prediction of a cut-off in the observed spectrum above a few $10^{19}$ eV due to interactions of UHE protons or nuclei with photons of the Cosmic Microwave Background (CMB). This prediction \cite{G66,ZK66} of the so-called GZK cut-off  (named after the authors of the original studies Greisen, Zatsepin and Kuzmin) was made in 1966, closely following the discovery of the CMB. This prediction started a long series of studies on the extragalactic propagation of UHECRs, including the production of secondary neutrinos and photons. 

In this paper, we review the main aspects of the extragalactic propagation of UHECR protons and nuclei. It will be organized as follows. In the next section, we will review the main interaction channels of protons and nuclei and discuss their influence on the energy and mass losses. In Section~3., we show some {\it propagated spectra}, allowing us to discuss  expectations concerning the evolution of the composition from the source to the Earth and the production of secondary protons. We discuss, in particular, the possible implications of the composition trend suggested by the recent analyses of the Pierre Auger Observatory. In Section.~4 we  discuss the potential effect of cosmic magnetic fields and, in particular, the possible departure from the rectilinear case. We then discuss the production of secondary messengers (neutrinos and photons) and the possible constraints their observation could bring for the understanding of the UHECR origin in Section.~5.  We finally conclude by briefly discussing the prospects for improving  our understanding of the UHECR phenomenon and the expected contributions of current and planed experiments in cosmic-rays, gamma-rays and neutrinos.

\section{Interactions of protons and nuclei with extragalactic photon backgrounds}

One of the most important aspects of the extragalactic propagation of UHECRs is the modeling of their energy and mass losses due to interactions. In the extragalactic medium, except maybe in the immediate vicinity of the UHECRs accelerators, only photon backgrounds are relevant. Besides CMB photons which represent the densest photon background, protons and nuclei interact mainly with infrared, optical and ultra-violet photons (hereafter IR/Opt/UV photons). In the past years, the models of IR/Opt/UV density and their cosmological evolution have been proposed for instance in \cite{Kneiske2004,Stecker2006}. These models are in good agreement in the far-infrared region at low redshift which is the most important for UHECRs propagation but differ significantly in the optical and UV range where the modeling of Kneiske and collaborators seems to better fulfill the constraints brought by the observation of distant blazars by Fermi \cite{FermiIR}. The difference between the different models only plays an important role in the production of cosmogenic PeV neutrinos but has a very limited impact on UHECR nuclei propagation (see below)\footnote{In the following we will mostly use the latest available update on the IR/Opt/UV background estimated by Kneiske and collaborators.}. 

\subsection{Interactions of protons} 
Besides the adiabatic losses due to the expansion of the universe , UHECR protons mainly suffer from the pair production mechanism, whose energy threshold with CMB photons is around $10^{18}$ eV, and  pion production, which is extremely efficient above $\sim7\times10^{19}$ eV. As mentioned earlier, these energy loss mechanisms and their relevance for the propagation of UHECR protons were pointed out very early after the discovery of the CMB photons\cite{PW65}. Greisen \cite{G66} and independently Zatsepin and Kuzmin \cite{ZK66}, estimated the opacity of the universe for cosmic-ray protons above $10^{20}$ eV due to the pion production mechanism and predicted the existence of the so-called GZK cut-off. A few months later Hillas \cite{Hillas67} and then Blumenthal \cite{Blumenthal70} studied the possible effect of pair production on extragalactic protons above $10^{18}$ eV while Stecker \cite{Stecker68} proposed a more sophisticated treatment of  photomeson interactions. The presence of a secondary flux of {\it cosmogenic} neutrinos as a consequence of the decay of photoproduced  charged pion was  shown by Berezinsky and Zatsepin \cite{BereOriginal} and the case of secondary cosmogenic gamma-rays and their possible contribution to the diffuse gamma-ray background was first studied a few years later by Strong, Wolfendale and Wdowczyk \cite{Strong73}. These early studies already contain most of the topics related to UHECR propagation that are currently studied, although more recent calculations can of course rely on better astrophysical inputs and datasets as well as larger computing capabilities. Concerning UHECR proton propagation, one of the important recent improvements was brought by Mucke and collaborators \cite{Mucke} who provided the monte-carlo generator SOPHIA for photomeson interaction of protons, based on an extensive study of the available data and phenomenological models, allowing a better handle on both the inelasticity of the interactions as well as the yield of secondary particles. Most recent calculations use the SOPHIA event generator.  

The energy evolution of the UHECR proton loss length, $\rm \chi_{loss}(E)$, is shown in Fig.~\ref{fig:crossmfp}a. The contribution of the different loss processes and the different photon backgrounds can be seen. The contribution of the IR/Opt/UV for pair production and pion production interaction is almost irrelevant on the whole energy range\footnote{This contribution is, however, important for the production of PeV neutrinos}. At low energy, up to a few $10^{18}$ eV, energy losses are dominated by the  contribution from the expansion of the universe\footnote{Let us note that the loss lengths are shown at z=0, the energy of the transition between expansion and pair production dominated losses actually depends on the redshift.}. The pair production mechanism starts to dominate just above its energy threshold. Finally, the pion production mechanism takes over above $\sim7\times10^{19}$ eV. Above $10^{20}$ eV the loss length drops to very low values, and the large scale universe becomes opaque to cosmic-rays, as pointed out in the original papers predicting the existence of the GZK cut-off. As we discuss later, this energy evolution of $\rm \chi_{loss}$  is unique to protons  and has extremely interesting consequences on the expected shape of the diffuse spectrum produced by a cosmological distribution of sources.

\subsection{Interactions of nuclei}

It is clear that the propagation of UHECR nuclei has been much less studied than the proton case. It is important to note that giant dipole resonance (GDR) interactions of nuclei were, however, explicitly mentioned in the original studies of \cite{G66,ZK66} and that the prediction of the GZK cut-off was not only for protons but also included complex nuclei.  The interactions of nuclei were later studied in more details by Stecker \cite{Stecker69}, and several studies discussed the potential contribution of nuclei to UHECRs at the $\rm14^{th}$ international cosmic-ray conference \cite{BereICRC75, Hillas75, Tka75, Puget75}. The most significant early study of UHECR nuclei propagation was certainly made by Puget, Stecker and Bredekamp \cite{PSB} who treated in great detail the propagation of nuclei ($\rm A\leq56$) and provided simple parametrizations based on the available data for the GDR and quasi-deuteron (QD) cross sections. Many years later, taking advantage of the improved constraints on the infrared background,  Stecker \cite{Stecker1998} and Epele and Roulet \cite{Epele} discussed the possibility, for the highest energy events detected at the time, to be heavy nuclei. Stecker and Salamon \cite{Stecker1999} proposed  a few months later a new calculation taking a better account of the GDR interaction energy threshold. The possibility of the presence of nuclei heavier than iron was furthermore studied by Anchordoqui et al. \cite{Ancho2001} while Bertone et al. \cite{Bertone} considered the effect of magnetic fields on the propagation of heavy nuclei and tested some astrophysical models for the origin of UHECRs. More recently, Khan et al \cite{Khan2005}, proposed new estimates of the GDR cross section, based on theoretical calculation using the TALYS nuclear reaction code \cite{TAL,Gor}, showing a better agreement with the available data than previous parametrizations from \cite{PSB}. In recent years the interest in the propagation of UHECR nuclei has significantly grown \cite{Allard05a, Arm2005, Sigl2004, Sigl2005, Hooper2007, Hooper2008, Ari2007, Aloi2008, Medi2009}. This is all the more true since the recent composition analyses of the Pierre Auger Observatory  seem to indicate a possible significant contribution of nuclei at the highest energies \cite{Augerer2010}. We will discuss some aspects of these recent studies in the next sections.  

 \begin{figure}[t]
\includegraphics[width=0.32\columnwidth]{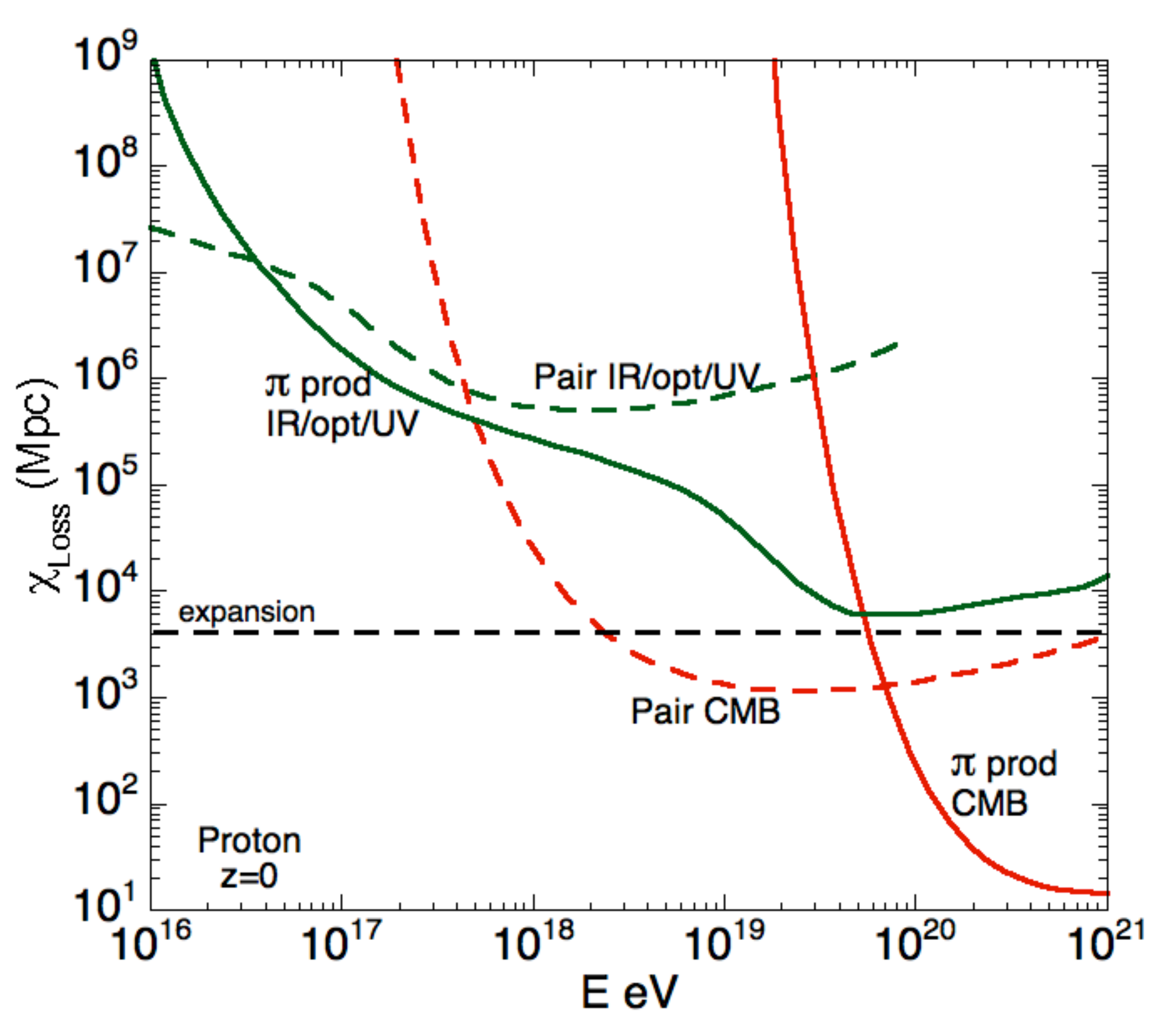}
\hfill
\includegraphics[width=0.32\columnwidth]{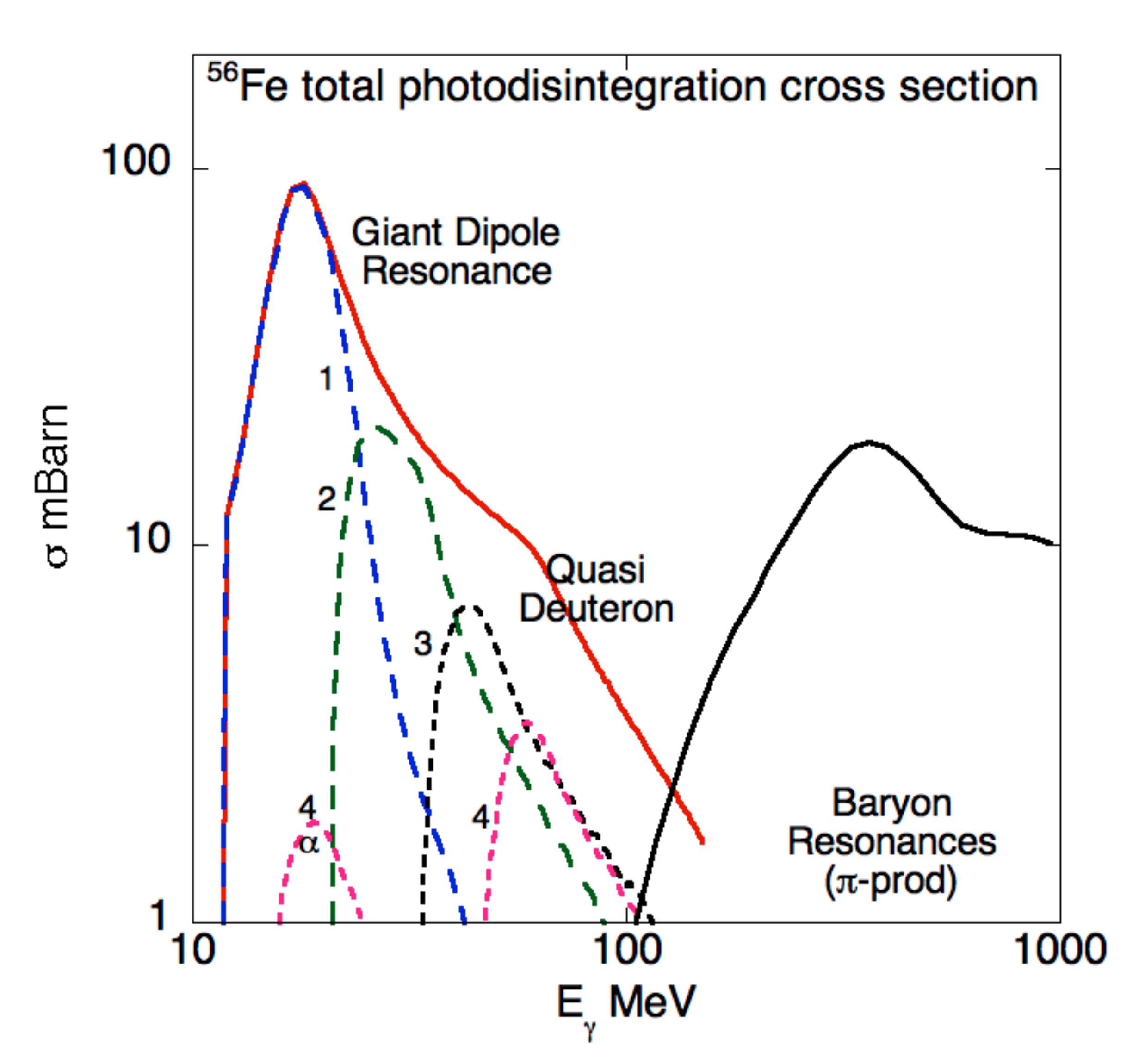}
\hfill
\includegraphics[width=0.32\columnwidth]{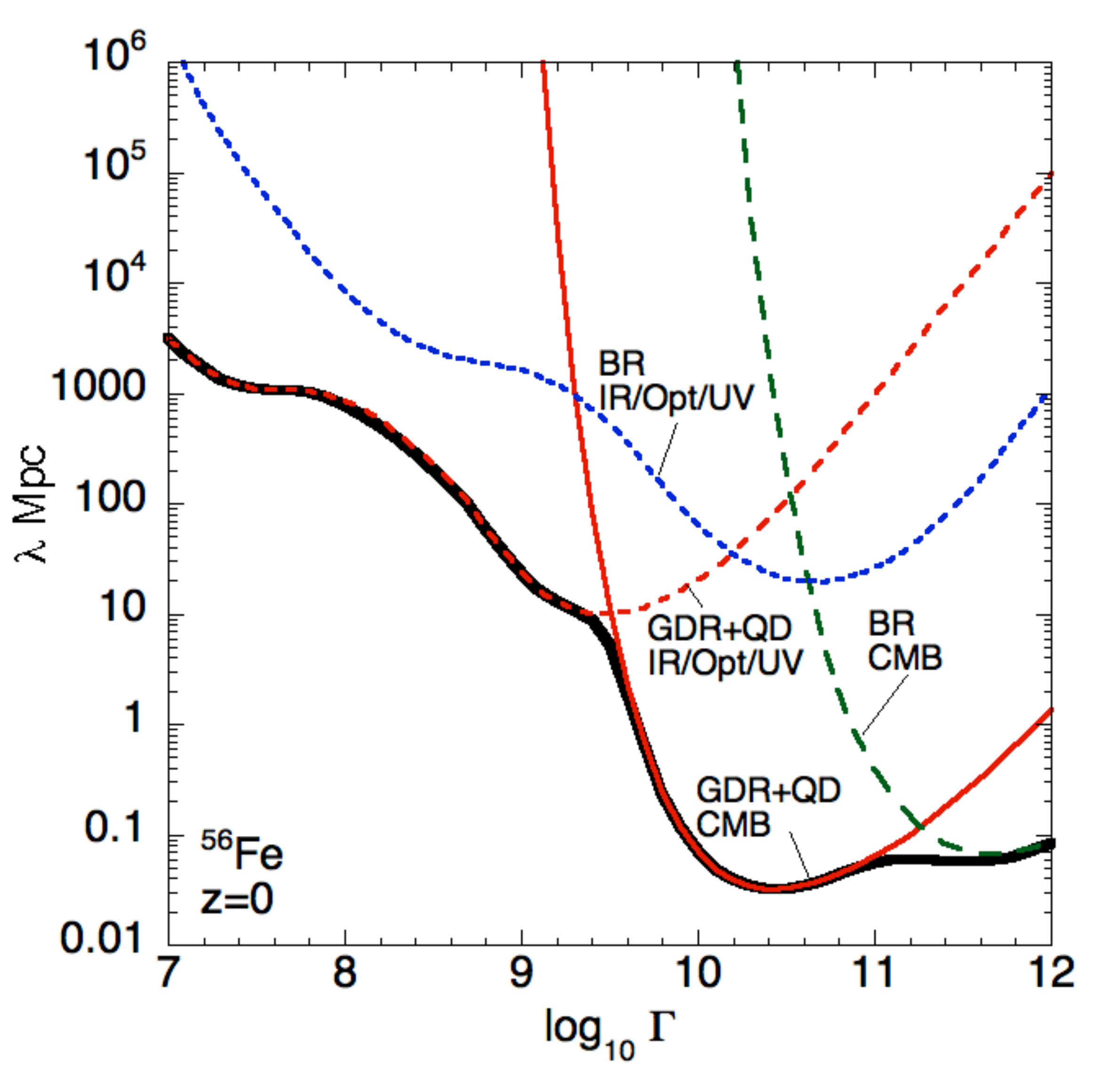}
\caption{Left : Energy evolution of the energy loss length, $\rm \chi_{loss}$, of protons, the contributions of different energy loss processes (adiabatic expansion, pair production and pion production) are displayed, as well as the different photon backgrounds (see labels). Center: Energy evolution of the photodisintegration cross section for $\rm^{56}Fe$, the contributions of the giant dipole resonance (GDR), quasi-deuteron (QD) and  baryon resonances (BR) are shown as well as the contribution of different nucleon multiplicities (for GDR and QD). Right: Lorentz factor evolution of the iron nucleus mean free path for the different photodisintegration processes and interactions with the CMB and IR/Opt/UV photons at z = 0. }
\label{fig:crossmfp}
\end{figure}

The interactions experienced by nuclei with photon backgrounds are different from the proton case. The most complete discussion of these interaction processes can be found in \cite{Rachen}. Unlike the proton case, one has to distinguish two categories of energy loss processes, those triggering a decrease of the nucleus Lorentz factor and those leading to the photodisintegration of the nucleus.  In the first category, nuclei propagation is mainly affected by adiabatic expansion and the pair production mechanism. For the latter, the energy threshold is proportional to the mass ,A\footnote{In the laboratory frame, for a given photon spectrum.}, of the parent nucleus whereas the loss length decreases like $\rm \sim A/Z^2$ at a given Lorentz factor \cite{Blumenthal70,Rachen}. 

 \begin{figure}[t]
\includegraphics[width=0.32\columnwidth]{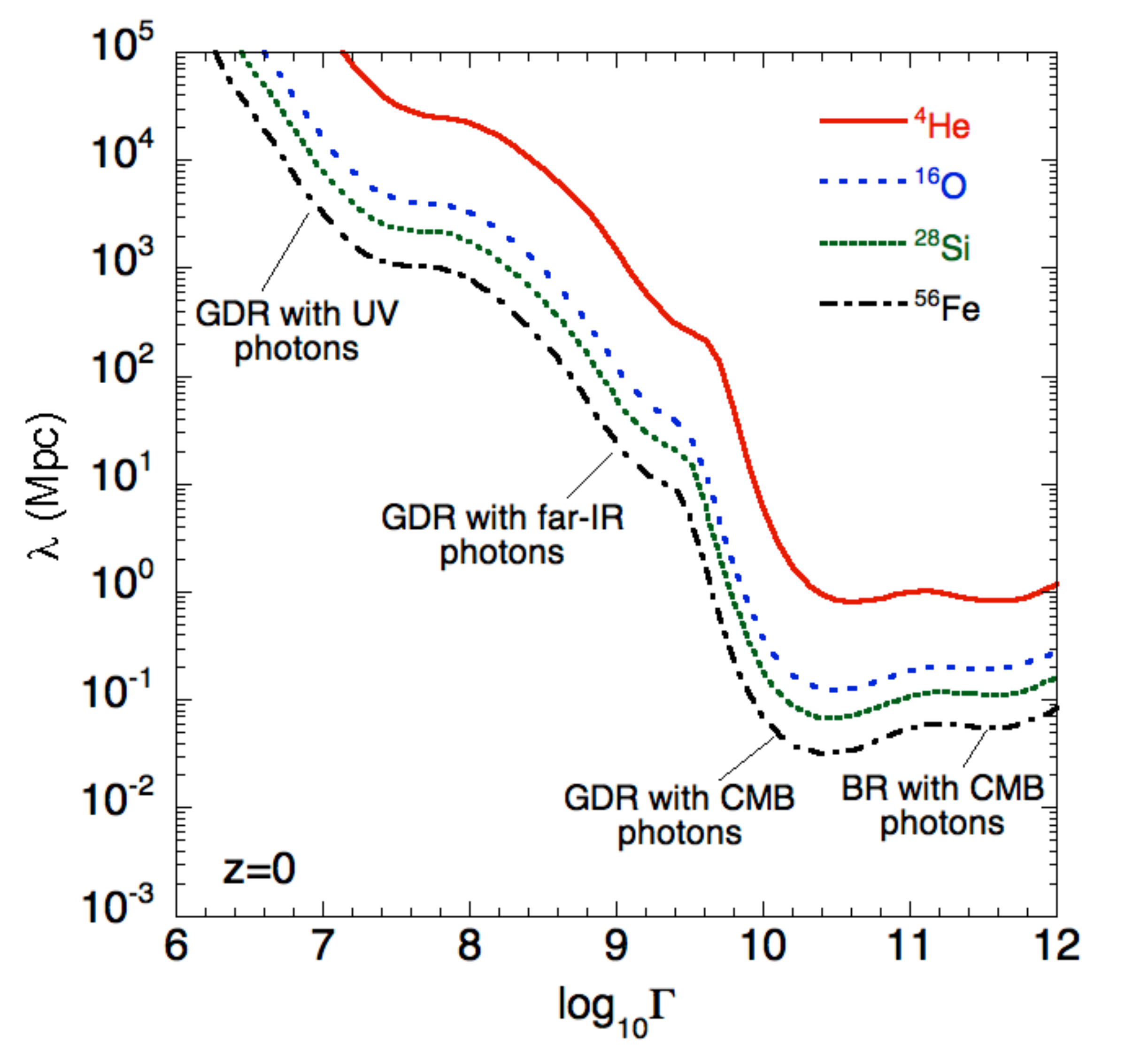}
\hfill
\includegraphics[width=0.32\columnwidth]{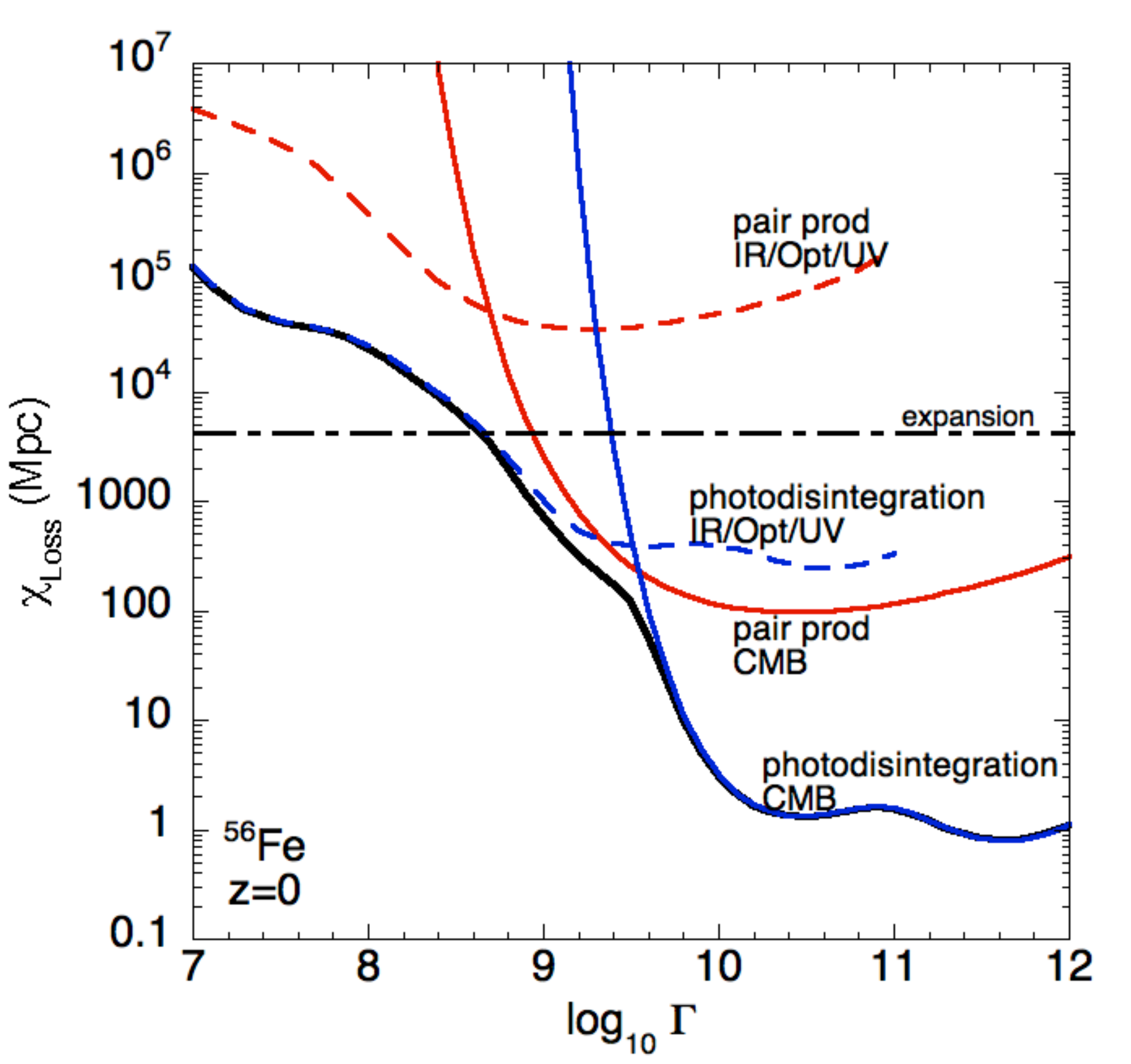}
\hfill
\includegraphics[width=0.32\columnwidth]{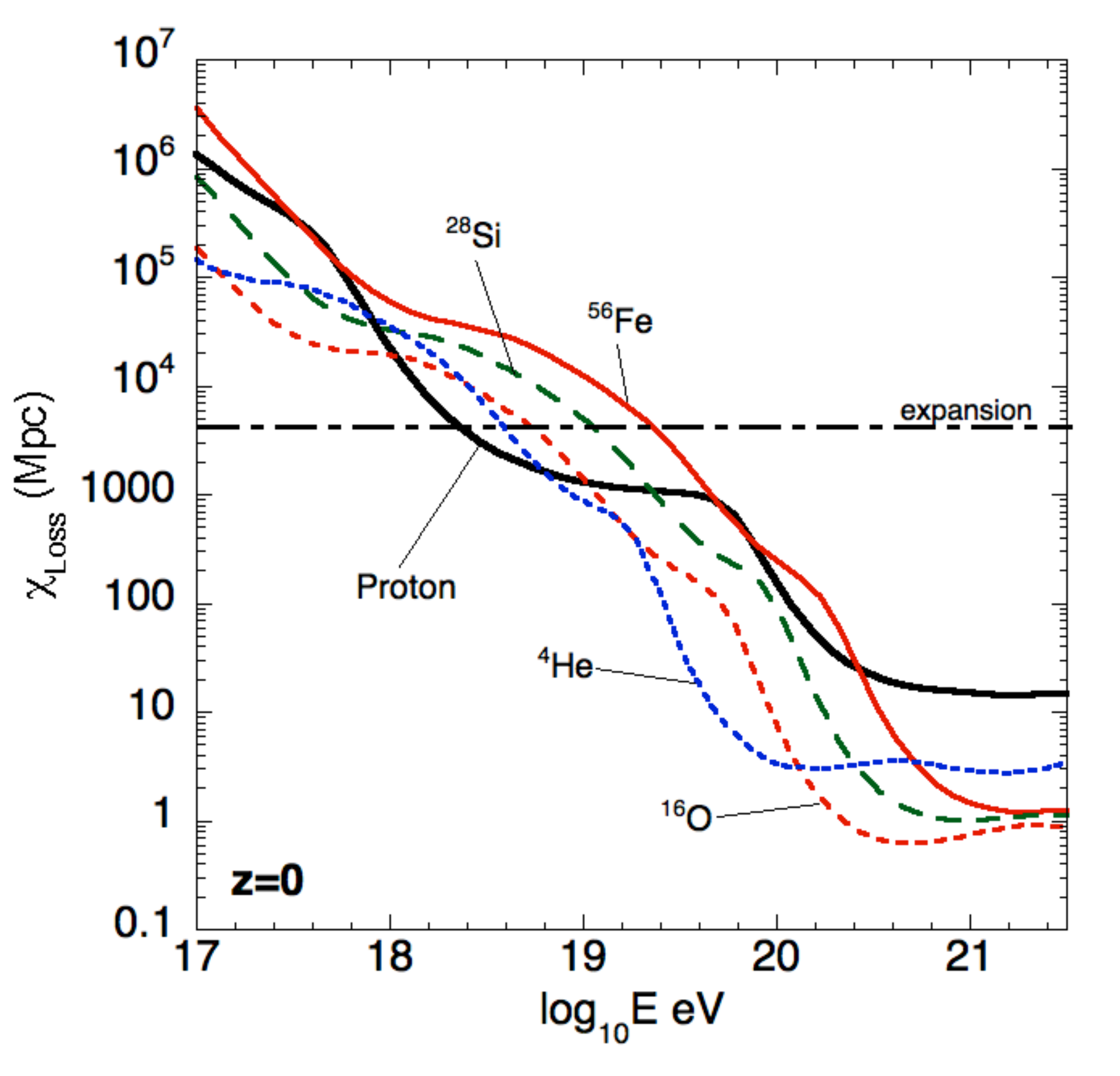}
\caption{Left: Total photodisintegration mean free path for various species (see labels) at z = 0. Center:  Evolution of the attenuation length of iron as a function of the energy at z = 0. The contribution of pair production and photodisintegration processes off the CMB and IR/Opt/UV photons are separated. Right: Comparison of the attenuation length of different nuclei at z = 0.}
\label{fig:lossnuclei}
\end{figure}

Concerning photodisintegration, different processes become dominant at different energies. The lowest energy and highest cross section process is the giant dipole resonance(GDR). The GDR is a collective excitation of the nucleus \cite{Chomaz} in response to electromagnetic radiation between $\sim$10 and 50 MeV\footnote{The threshold for most nuclei is between 10 and 20 MeV except for peculiar cases like $\rm ^9Be$ or the dinucleon and trinucleon} where a strong resonance can be seen in the photoabsorption cross section (see Fig.~\ref{fig:crossmfp}). The GDR mostly triggers the emission of one nucleon (most of the time a neutron but depending on the structure of the parent nucleus, $\rm \alpha$ emission can also be strong for some nuclei), 2, 3 and 4 nucleon channels can also contribute significantly though their energy threshold is higher. Around 30 MeV in the nucleus rest frame and up to the photopion production threshold, the quasi-deuteron (QD) process becomes comparable to the GDR (but much lower than at the peak of the resonance) and its contribution dominates the total cross section at higher energies.  Unlike GDR, QD is predominantly a multi-nucleon emission process. The photopion production (or baryonic resonances (BR)) of nuclei becomes relevant  above 150 MeV in the nuclei rest frame  (e.g.,  $\sim5\times 10^{21}$ eV in the lab frame for iron nuclei interacting with the CMB). The large excitation energy usually triggers the emission of several nucleons in addition to a pion that might be re-absorbed before leaving the nucleus. Below 1 GeV, the cross section is in good approximation proportional to the mass, the reference being the deuteron photoabsorption cross section \cite{Rachen}. Above 1 GeV, nuclear shadowing effect become important and the scaling in A of the cross section is no longer valid. Furthermore, the fragmentation of the nuclei into small pieces starts to compete with the evaporation of a few nucleons \cite{Rachen}. The GeV energy range  in the center of mass of a nucleus correspond to laboratory frame energies above  well above $10^{21}$ eV even for light nuclei and can then be safely omitted in propagation calculations \footnote{This process would take place at lower laboratory frame energies with IR/Opt/UV photons but, in this case, it is totally screened by lower energy interactions (GDR) with CMB photons.}. The photodisintegration cross section for $\rm ^{56}Fe$ nuclei (for photons between 1 MeV and 1 GeV in the nucleus rest frame) and the contribution of the different processes can be seen on Fig.~\ref{fig:crossmfp}a. Let us note that the energy evolution of the yield (e.g, the cross section multiplied by the number of emitted nucleons) is flatter due to the contribution of multi nucleon emission. In the GDR energy range, the hierarchy between the different channels can highly vary from one nucleus to another, some nuclei, for instance, being more likely to emit a proton or sometimes an $\rm \alpha$ particle depending on the detail of its nuclear structure \cite{Khan2005}. 

The corresponding mean free path for photodisintegration interactions with CMB and IR/Opt/UV for $\rm ^{56}Fe$ is displayed in the right panel of Fig.~\ref{fig:crossmfp}. One can see the dominant role played by GDR interactions first with IR/Opt/UV photons and then with the CMB. For very large values of the Lorentz factor, above $\rm\Gamma=10^{11}$, the baryonic resonances take over and prevent the mean free path from recovering to larger values when the influence of the GDR and QD decreases.  At the peak of the influence of the GDR (around $\rm \Gamma=10^{10.5}$) the mean free path for Fe nuclei can be as low as $\sim30-40$ kpc, e.g much lower than the mean free path of protons for photopion interactions.  Fig.~\ref{fig:lossnuclei}a shows a comparison of the total mean free path for different nuclei. The several features visible in the energy evolution of the mean free path (see labels and \cite{Denis2008} for more details)  take place at similar Lorentz factors for all nuclei (which means at energies approximately proportional to the mass A of the nucleus), due to the slow scaling with the mass of the GDR threshold in the nucleus rest frame (the difference between $\rm^4$He ($\sim$ 20 MeV) and $\rm^{56}$Fe ($\sim$12 MeV) is however visible). On the other hand, the scaling of the mean free path $\rm\lambda$ with A reflects the scaling of the interaction cross section.

\begin{figure}[t]
\includegraphics[width=0.32\columnwidth]{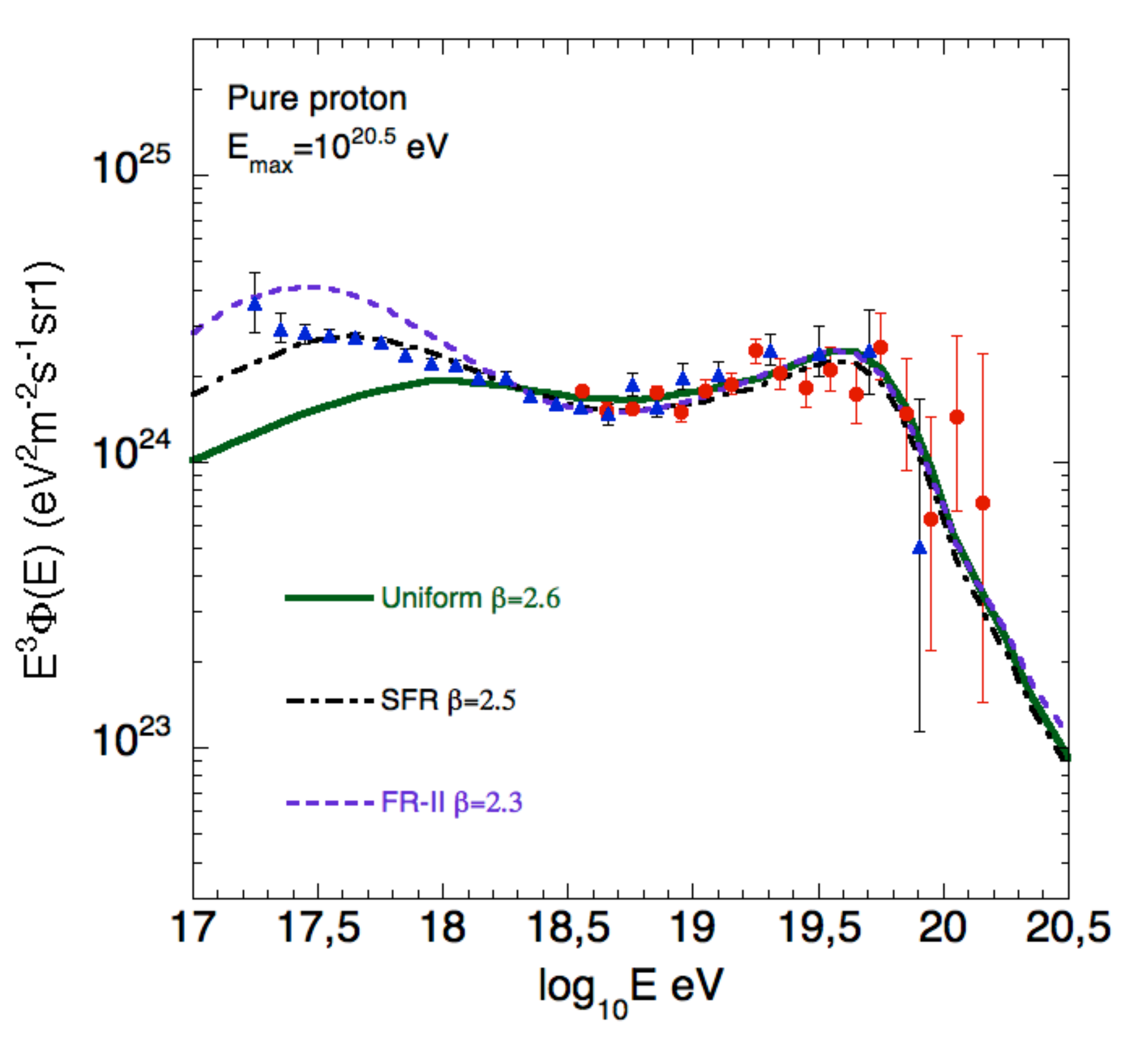}
\hfill
\includegraphics[width=0.32\columnwidth]{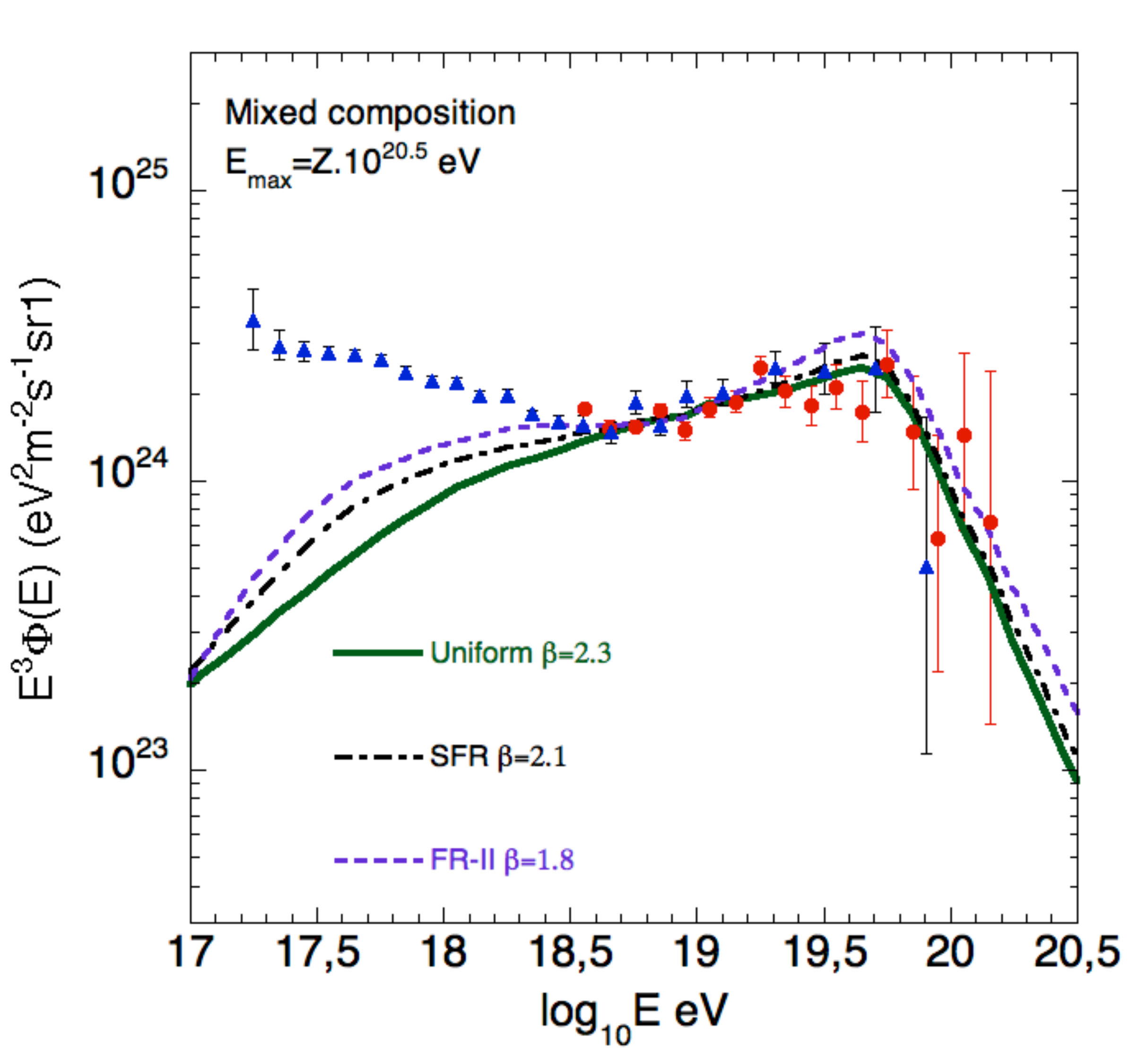}
\hfill
\includegraphics[width=0.32\columnwidth]{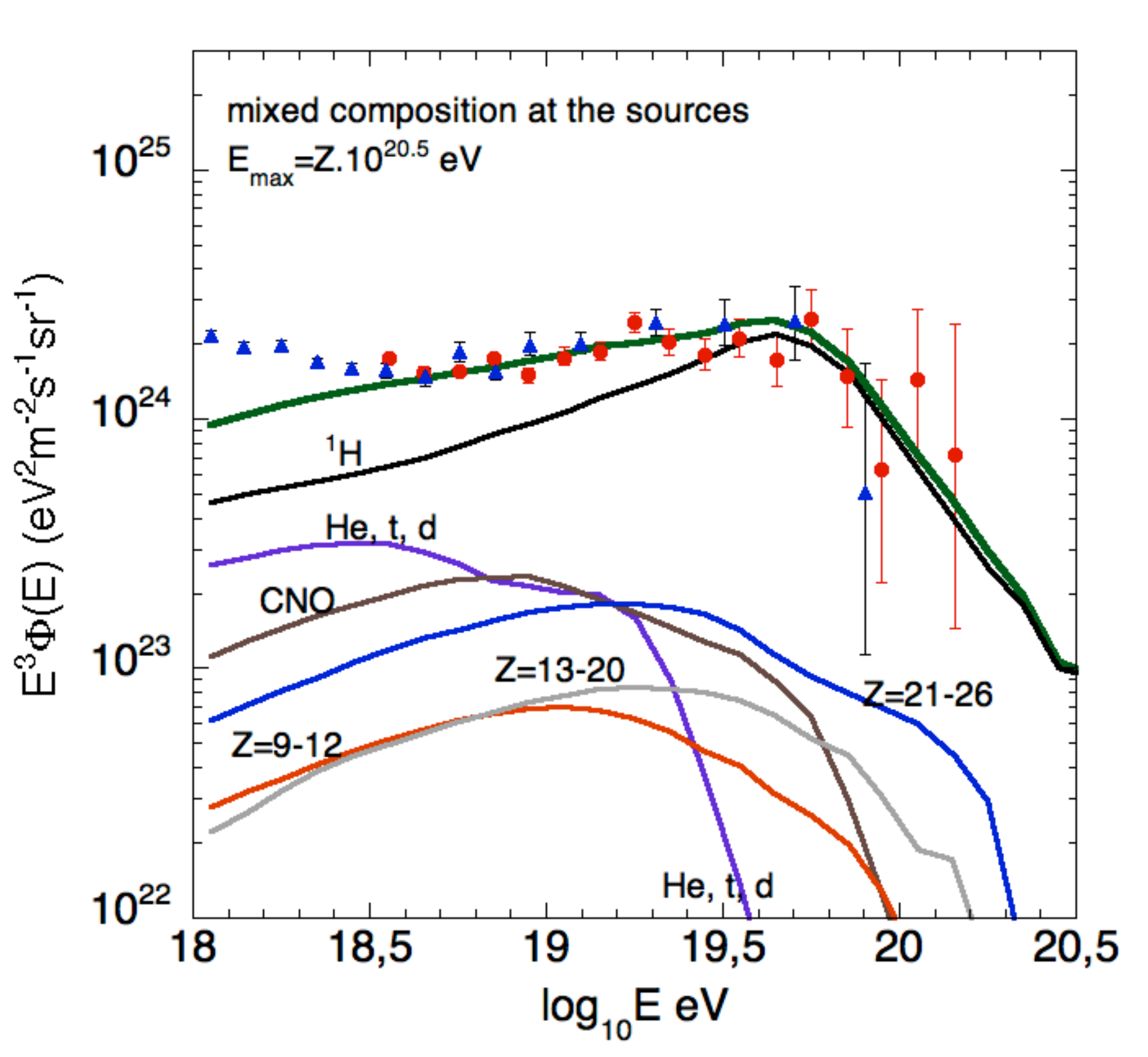}

\caption{Left: Propagated spectra assuming a pure proton composition at the extragalactic sources for different hypotheses on the cosmological evolution of the cosmic-ray luminosity (see legend), the maximum energy at the sources is $\rm E_{max}=10^{20.5}$ eV. The spectra are compared with HiRes monocular data. Center:  Same as left panel but assuming a mixed composition at the sources similar to low energy galactic cosmic-rays, the maximum energy of the different elements scale proportionally with the charge : $\rm E_{max}(Z)=Z\times10^{20.5}$ eV. Right : Same as the central panel, but the different elemental groups contribution is shown (for the "uniform" evolution case in the central panel).}
\label{fig:dipmix}
\end{figure}

Fig~.\ref{fig:lossnuclei}b shows the contribution of pair production and photodisintegration processes to the total attenuation length of iron nuclei. Above $\rm\Gamma\sim10^{8.8}$ photodisintegration processes take over and dominate the energy losses through most of the energy range. The pair production effect is smaller and only dominant, at low redshift, for Lorentz factors of the order of $\rm\Gamma\sim10^{9.5}$. Although the competition between pair production off the CMB and photodisintegration processes with IR/Opt/UV photons depends on the redshift (e.g. when the redshift increases the contribution of pair production is larger due to the stronger evolution of the CMB), the energy losses of nuclei are mainly dominated by photodisintegration processes. A comparison between the attenuation lengths of different species is displayed in Fig~.\ref{fig:lossnuclei}c. The difference in the shape of the evolution of $\rm\chi_{loss}$ between the proton and complex nuclei cases is clearly visible (we will discuss the consequences in the next section). For complex nuclei, two sharp drops are visible in the energy evolution of $\rm\chi_{loss}$, taking place at energies more or less proportional to the mass of the parent nucleus. The first one (around $\rm\sim A\times 10^{18}$ eV at z=0) is due to the combined effect of (mainly) photodisintegration on far-infrared photons and pair production on CMB photons and the second (even sharper, around $\rm\sim A\times 4\,10^{18}$ eV at z=0) to photodisintegration on CMB photons. These features in the loss length are expected to be seen in the propagated spectra (see below) for the different elemental groups. Besides the shape of the $\rm\chi_{loss}$ evolution, one should note that one of the major differences between protons and complex nuclei propagations is that, unlike protons,  a given nucleus does not remain on the same attenuation length curveÕ during its propagation as a result of the dominance of mass losses over  Lorentz factor decrease through most of the energy range (at least at low redshift). In particular, as a result of the very fast photodisintegration with CMB photons, elemental groups should simply disappear from the spectrum at high energy one after the other instead of pilling up at lower energy as in the proton case (see for instance \cite{Denis2008} for a more complete discussion). Let us note that these features in the evolution of $\rm \chi_{loss}$, and especially the very low values reached in the CMB regime, were already pointed out in some of the early studies mentioned above. The possibility of detailed predictions was however limited at that time, partly due to the poor constraints on the IR/Opt/UV backgrounds.

\section{Propagated spectra and evolution of the composition}

Once the energy losses of UHECR proton and nuclei are characterized one can in principle calculate propagated spectra by assuming a cosmological distribution of UHECR sources, including the evolution of their luminosity as well as their injection spectra. The sources are most of the time assumed to be standard candles (e.g, same luminosity at a given redshift, same injection spectral index and same maximum injection energy or rigidity), for the sake of simplicity. We will limit ourselves to these hypotheses in this section but discuss the potential impact of adding astrophysical free parameters.

\subsection{pure proton case}

We start our discussion by assuming a pure proton composition at the extragalactic sources. Fig.~\ref{fig:dipmix}a shows propagated spectra assuming different cosmological evolutions for the cosmic-ray luminosity (namely, no evolution labeled as "uniform", star formation rate (SFR) type evolution \cite{Hopkins} and an evolution similar to that of strong radio sources, hereafter labeled as FR-II \cite{Wall}) adjusted to the HiRes monocular spectra \cite{Cut}. The features that can be seen seen in the spectrum are a direct translation of the $\rm\chi_{loss}$ evolution discussed in the previous section : at low energy, energy losses are dominated by the adiabatic expansion (independent of the energy) and the spectrum more or less keeps the "injected shape." Above the pair production threshold, $\rm\chi_{loss}$ decreases quite abruptly with the energy which triggers a steepening (softening) of the propagated spectra (this feature is called the "pair production bump in \cite{Bere88}). Well above the pair production threshold, the evolution of  $\rm\chi_{loss}$ flattens, which results in a hardening of the spectrum. The sequence is called the "pair production dip." Above the pion  production threshold, 
 $\rm\chi_{loss}$ decreases again abruptly to reach very low values and the propagated spectrum experiences a very steep softening which is the so-called GZK cut-off for protons. This sequence of features in the expected proton propagated spectrum was understood very early in the pioneering studies (see the previous section) and studied in more detail in  \cite{Hill85} and \cite{Bere88}. Let us note that the position and amplitude of the different features depend on the cosmological evolution of the CR luminosity (hereafter the cosmological evolution for short). Indeed, the actual value of $\rm\chi_{loss}$ depends on the energy and density of the photon background which evolves with redshift. If one assumes a cosmological evolution which gives a larger weight to distant sources, the transition between adiabatic and pair production losses will take place at lower energy and the pair production dip will be deeper and slightly shifted to lower energies. As a consequence, the spectral index at the sources (assuming standard candles) required to fit the experimental data also depends on the cosmological evolution of the sources. 
 
 One of the remarkable characteristics of the pure proton case is its ability to reproduce the ankle of the cosmic-ray spectrum with only the extragalactic component. This observed spectral feature is in this case associated with the energy losses of extragalactic protons and not with the transition from galactic to extragalactic cosmic-rays (hereafter the GCR to EGCR transition) as proposed in most models. The "pair production dip" as an astrophysical interpretation for the ankle of the cosmic-ray spectrum as well as its implications  for the GCR to EGCR transition (which has to take place at lower energy) have been proposed by Berezinsky and collaborators \cite{BereAstro, Bere2004, Bere2005a, Bere2005, Bere2006, Bere2008} and discussed by  other authors (see for instance \cite{DDM2005, KS2006}). 
 
 Let us note that  a few additional parameters need to be added to the dip model (compared to what is shown in Fig~.\ref{fig:dipmix}a).The spectral indices required to correctly fit experimental spectra can be relatively soft (2.6-2.7 for the uniform evolution case and 2.5 for the SFR) and as a result the energy budget required at the sources down to GeV can rapidly become prohibitive. To avoid this problem a change in the spectral index (from 2.0 below $\sim10^{18}$ eV to 2.6-2.7 above) due to acceleration mechanism at the sources (supposed to be AGNs) was proposed in \cite{BereAstro}. Moreover \cite{KS2006} obtained a quantitatively equivalent effect by showing that the soft spectral index could simply be due to the convolution of the spectral index at the sources,  $\rm\beta_{source}$ (2.0, "in agreement" with simplest scenarios of diffusive shock acceleration), with the distribution function of the maximum energy at the sources  ($\rm dN(E_{max})/dE_{max} \propto E_{max}^{-\alpha}$ with $\rm\alpha=1.6-1.7$) leading to an effective spectral index $\rm\beta_{eff}=\beta_{sources}+\alpha-1=2.6-2.7$. These additional "low energy cut mechanisms" also lead to a reduction of the expected proton fraction at $10^{17}$ eV that might otherwise overshoot observations in the unmodified version of the model. Magnetic horizon effects at low energy (see below) have also been invoked to reduce the proton fraction at low energy. This low energy cut mechanism would however leave the energy budget unchanged when compared with the "default" scenario.
 
 \subsection{mixed and nuclei dominated compositions}
 
 One of the consequences of adding complex nuclei in the UHECR source composition is to impact the prediction of the presence of a pair production dip. Indeed, as mentioned in the previous sections, the pair production dip is a signature of extragalactic proton energy losses and is "encoded" in the energy evolution of the loss length. As the energy losses of complex nuclei and their energy evolution are different from the proton case, one does not expect such a feature in their propagated spectra. One can  show  that the possibility of observing a pair production dip in the extragalactic cosmic-ray spectrum requires the abundance of complex nuclei at the source to be less than 10-20\% (depending on the mixture of complex nuclei assumed) \cite{Bere2005a, Allard05a, Denis2007a}. 
 
 As the extragalactic source composition is extremely difficult to predict, as the UHECR sources are still unknown, most of the first studies on the extragalactic propagation of nuclei considered the cases of pure compositions of complex nuclei and especially pure iron, this nucleus being likely to contribute at the highest energies due to its high energy threshold for photodisintegration with CMB photons. \cite{Allard05a} proposed to study the case of a mixed composition equivalent to that of low energy galactic cosmic-rays as a possible composition (see also \cite{DenisNeut2006} for more details). Even though there is no strong reason for the composition at the extragalactic source to be exactly the same as that of low energy galactic cosmic-rays (which moreover is not known extremely precisely, as it has to be reconstructed from the observations \cite{Vernois}), the hierarchy of the relative abundances between the different elements could be somewhat preserved. Mixed compositions are, moreover, a good case to study the energy range in which the different species are likely to contribute. The propagated spectra for a mixed composition, compared to the monocular spectra from HiRes are represented in Fig.~\ref{fig:dipmix}b. As discussed earlier, the presence of a pair production dip is not expected in this case, and the experimental spectrum can only be reproduced down to the ankle with the extragalactic component, (this remains true even if the composition is dominated by complex nuclei) which is in this case the signature of the end of the GCR to EGCR transition. 
 
From the point of view of the GCR to EGCR transition, the mixed composition case is an intermediate between the dip (steep transition that ends before the ankle) and the classic ankle model (a late transition where the galactic component extends up to energies above $10^{19}$ eV, see for instance \cite{WW}). The main difference with the dip model (besides the interpretation of the ankle) comes from the smoother transition implied by the mixed composition model and the evolution of the composition above the ankle. In \cite{Denis2005b, Denis2007a, Denis2007} it was argued that these models (degenerated from the point of view of the UHECR spectrum) could be distinguished by the difference they imply in the evolution of the composition in the GCR to EGCR transition energy range or above. The authors calculated the expected evolution of the mean  depth of  the air shower maximum (hereafter $\rm \langle X_{\max} \rangle$) to extract expected experimental signatures of the different models and argued that the mixed composition model was the most compatible with experimental data available at the time (see the above mentioned reference for a larger discussion of the GCR to EGCR transition and the review by Aloisio and Berezinsky \cite{Beretrans}).

The contribution of the different elemental groups is displayed in Fig.~\ref{fig:dipmix}c, for the "uniform" evolution case. The maximum energy at the sources (above which an exponential cut-off is added) was assumed to be $\rm E_{max}=Z\times10^{20.5}$ eV, the same maximum energy for protons as in the previous case and proportional to the charge of the nucleus as expected in confinement limited acceleration processes. It means that all the species are assumed to be accelerated well beyond the "GZK energy". The source distribution is assumed to be continuous down to a distance of 4 Mpc from Earth (which is the distance of the closest AGNs such as Cen A). The proton component obviously presents the same features as described above. For complex nuclei, the behavior of the elemental spectra can also be easily understood from the energy evolution of $\rm\chi_{loss}$. The different groups show first  a smooth inflexion in their spectrum around $\rm\sim A\times 10^{18}$ eV (actually a bit lower for nuclei heavier than He) due to the combined effect of photodisintegration on IR photons and pair production with CMB photons, and  then a catastrophic drop above $\rm\sim A\times 5\,10^{18}$ eV due to photodisintegration with CMB photons. As a result the contribution of light and intermediate nuclei should decrease as the energy increases and then disappear above the interaction threshold with CMB photons. At $10^{20}$ eV for instance, besides protons, only nuclei  heavier than $\sim$ Si should significantly contribute. Above $\sim3\,10^{20}$ eV, the heavy nuclei should finally disappear (once more provided there is no source much closer than 4 Mpc from the Earth and no abundant nuclei significantly heavier than Fe) and only protons, (besides possibly small fragments produced close to the Earth and provided a strong enough accelerator is located at less than 30-50 Mpc) should be present in the cosmic ray flux (if any). To summarize, if the maximum energy per unit charge at the sources reaches large values (say above $10^{20}$ eV), one expects an initially mixed composition to become gradually lighter\footnote{With however an expected composition feature between $\sim$50 and 300 EeV, see \cite{Denis2007,Denis2008} and below.}  as a result of the photodisintegration of complex nuclei.

\begin{figure}[t]
\includegraphics[width=0.32\columnwidth]{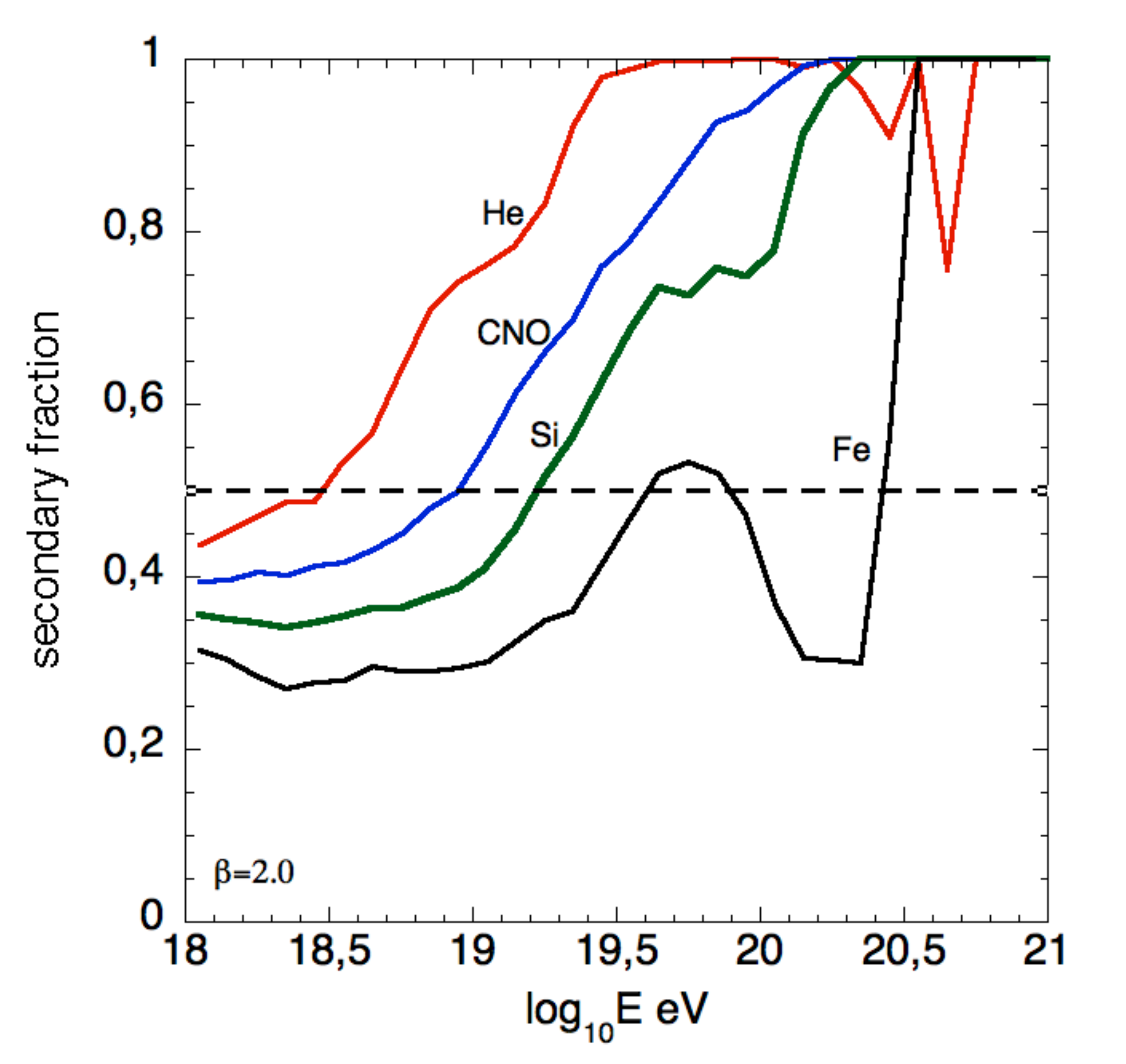}
\hfill
\includegraphics[width=0.32\columnwidth]{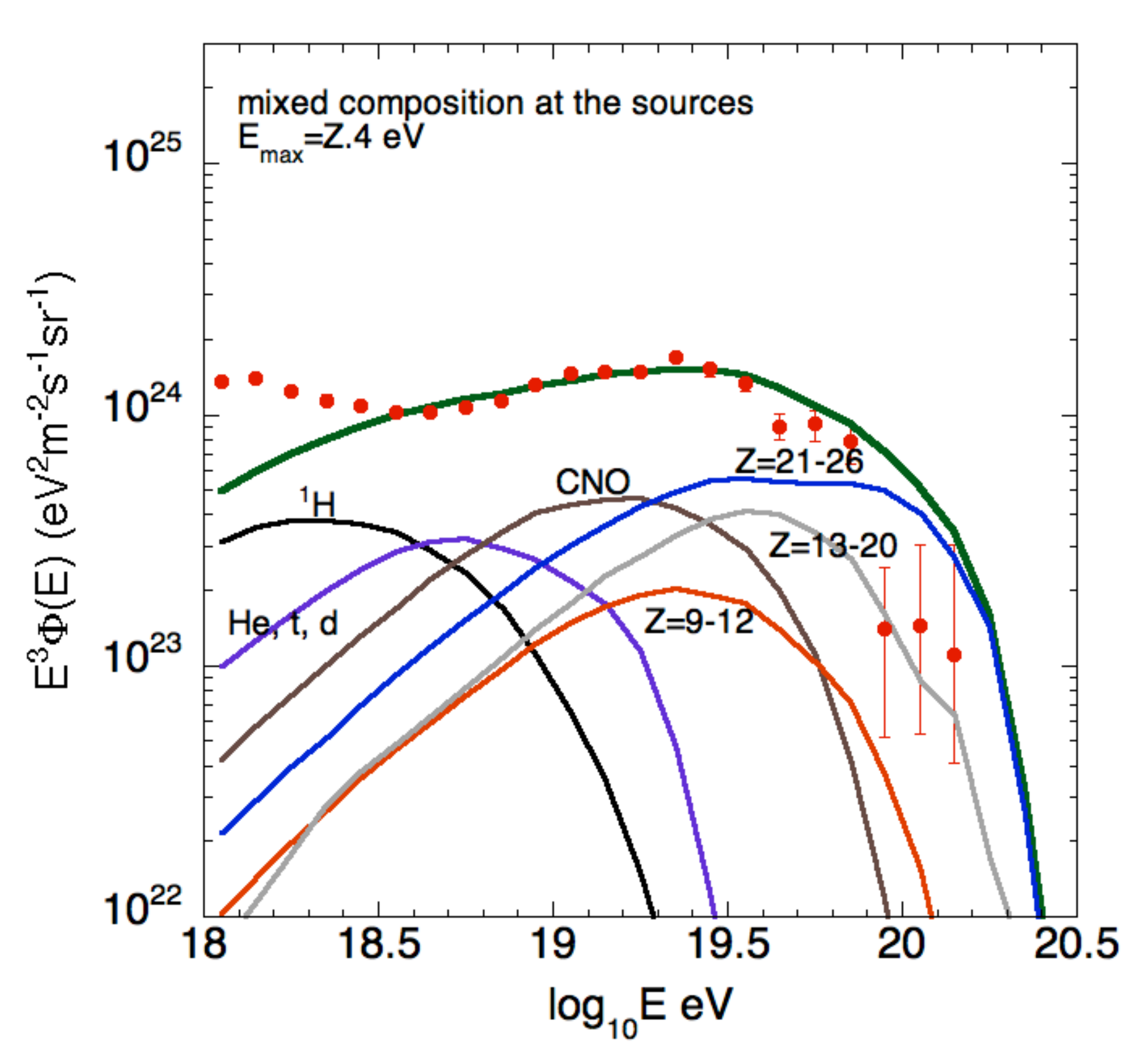}
\hfill
\includegraphics[width=0.32\columnwidth]{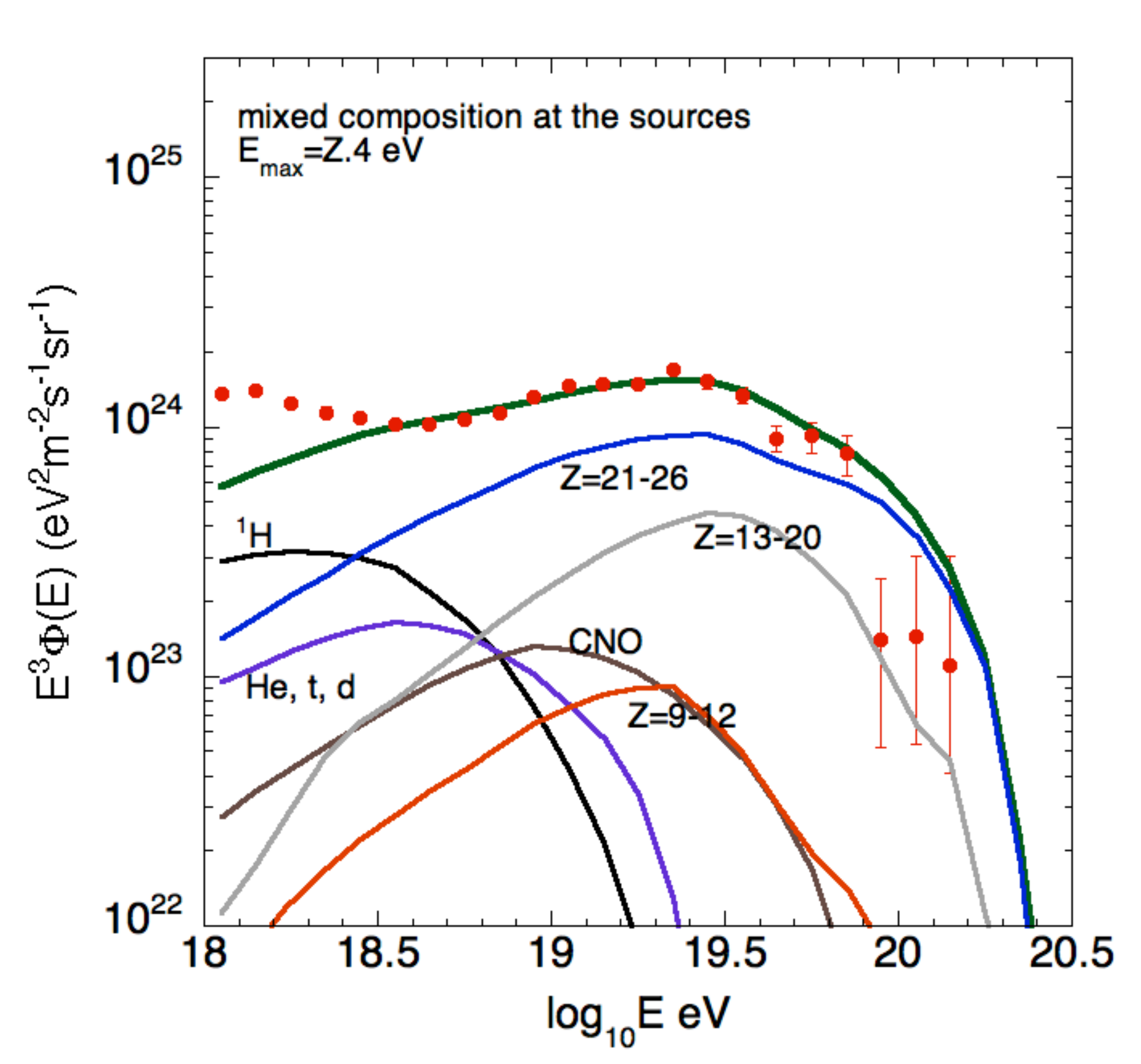}

\caption{Left: Relative abundance of secondary nucleons, dinucleons, trinucleons and $\rm\alpha$-particles in the propagated spectra assuming different pure complex nuclei composition at the sources (see labels), a source spectral index $\rm \beta=2.0$ and maximum energy $\rm E_{max}(Z)=Z\times10^{20.5}$ eV. Center : Propagated spectrum assuming the same mixed composition as in Fig.~\ref{fig:dipmix}b, the maximum energy at the sources is $\rm E_{max}(Z)=Z\times4\,10^{18}$ eV  and the spectral index $\rm \beta=1.6$. The propagated spectrum is compared to Auger data \cite{Augersp2011}. Right : Same as the central panel, but for a mixed composition enriched in heavy elements (30\% of the source composition), a maximum energy $\rm E_{max}(Z)=Z\times4\,10^{18}$ eV and a spectral index $\rm \beta=2.0$.}
\label{fig:lowemax}
\end{figure}

Let us now turn to nuclei dominated compositions and, in particular, pure nuclei compositions. In these cases, the light component in the extragalactic composition is provided by the emission of nucleons due to photodisintegration processes. Above an energy $\rm \sim A\times5\,10^{18}$ eV (depending on redshift) nuclei interact with CMB photons and are photodisintegrated both very rapidly \cite{PSB} and completely. Above $\sim 5\,10^{18}$ eV secondary nucleons (emitted by a primary of mass A and charge Z) are to good approximation injected "immediately" (this approximation holds only for reasonably distant sources) with the same spectral index as the primary nuclei up to an energy $\rm E_{max}(Z)/A$ and with a relative abundance $\rm A^{2-\beta}$ (where $\rm \beta$ is the source spectral index) compared to primary nuclei at the same energy. The photodisintegration of nuclei slows down as the energy decreases  and the injection of secondary nucleons is then harder than the primary nucleus spectral index (and much more spread in time).  The energy evolution of the composition is afterwards affected by the energy losses of the primary component and the secondary nucleons the same way as in the mixed composition case. This is illustrated on Fig.~\ref{fig:lowemax}a, where the energy evolution of the relative abundance of secondary fragments \footnote{In a vast majority nucleons but also dinucleons, trinucleons, and $\rm \alpha$ particles} is shown assuming the different cases of pure composition at the sources (He, CNO, Si and Fe), a spectral index $\rm \beta=2.0$, and, a maximum energy $\rm E_{max}(Z)=Z\times 10^{20.5}$ eV at the sources. A detailed discussion of the different features in the energy evolution of the secondaries abundance can be found in \cite{Denis2008}. We can note however that the secondaries abundance is initially below 50\% (which is the value expected with the $\rm A^{2-\beta}$ for $\rm \beta=2.0$) mainly due to the larger energy losses of protons. Once the proton component recovers from pair production losses (i.e, the evolution of $\rm\chi_{loss}$ flattens), the abundance of secondaries increases (also due to the decrease with the energy of $\chi_{loss}$ for nuclei), and reaches 50\% approximately when the secondary protons and the primary nuclei loss length become equivalent. A flattening or even a decrease can be seen if protons reach the pion production threshold with CMB photons before nuclei start to be photodisintegrated by the CMB (this feature can be quite noticeable for heavy nuclei, see Fig.~\ref{fig:lowemax}a). Ultimately at the photodisintegration threshold with CMB photons the primary component disappears, and the composition is totally dominated by light fragments (provided the maximum energy at the sources is large enough which is the case in our illustration). This feature takes place respectively above $\sim1.2\,10^{20}$, $2\,10^{20}$, and $3\,10^{20}$ eV for CNO, Si and Fe primaries (for the choice of 4 Mpc for the distance of the closest source we made in this example) \footnote{Let us note that the highest energy event ever recorded, by the Fly's Eye experiment \cite{Fly1, Fly2}, with a reported energy of $\rm(3.2\pm0.9)\times10^{20}$ eV could be a heavy nuclei without requiring an unrealistically close source. However, the source would have to be located within $\sim20$ Mpc, even at the lower bound of the confidence interval.}. This final feature has interesting implications : if one assumes that the composition is dominated for instance by heavy nuclei around $10^{20}$ eV (which is the case in our illustration in the Fe primary case) then the flux is expected to drop dramatically around $3\,10^{20}$ eV (or a bit earlier if the closest source is much more distant than 4 Mpc). The mean rigidity of the cosmic-ray above this energy (if there is any flux) is also expected to be much larger\footnote{Of course we need to assume here that nuclei well heavier than Fe have an insignificant contribution to the UHECRs.}. This makes  this energy range potentially interesting for cosmic-ray astronomy provided one is able to build an experiment with a large enough exposure, which should be the case for the expected JEM-EUSO experiment \cite{Medina}.
In any case, one can see that, even assuming pure nuclei compositions, if the acceleration maximum energy is large enough one expects the composition to become "globally" lighter with energy (with relative abundances depending on the source spectral index), and ultimately completely dominated by light secondaries after a sharp collapse of the primary component due to photodisintegration with CMB photons.

 \subsection{Impact of Auger composition analyses}  
 
 The Pierre Auger collaboration, recently provided the largest statistics composition analyses above the ankle based on the measurement of the mean maximum depth of shower development ( $\rm \langle X_{\max} \rangle$) and its associated dispersion \cite{Augerer2010, Augerer2011}. When compared with air shower simulations, the energy evolution of $\rm \langle X_{\max} \rangle$ and its dispersion (hereafter RMS)  strongly suggest that the composition, light or mixed at the ankle, is evolving toward heavier elements as the energy increases.  This trend is especially striking in the evolution of the RMS, which becomes very narrow and close to what is expected for heavy dominated compositions (and a very low abundance of protons) in the last energy bins around $10^{19.5}$ eV\footnote{It is important to mention that the interpretation of shower depth measurements in terms of cosmic-ray composition rely on the comparison with air-shower simulations and the modeling of hadronic interactions at energies well above what is measured in accelerators. It has to be kept in mind that the evolution of shower depth with energy could be due to some unexpected behavior of hadronic interactions at very high energy.  Current and future constraints from LHC on high energy hadronic phenomena taking place in air shower development \cite{Tanguy} will be crucial to estimate how likely this alternative interpretation of shower depth measurements is.}. The HiRes collaboration claims a light dominated composition in the whole energy range above the ankle \cite{HiResCompo}. The HiRes statistics above $10^{19}$ eV (where the HiRes collaboration claims a different behavior of the shower depth) are however $\sim$ three times lower than the current statistics provided by the Pierre Auger Observatory. It is then not obvious that this discrepancy is in fact statically significant (see \cite{Michi} for a review on recent composition analyses). This issue should be settled by the Telescope Array experiment which, although smaller than the Pierre Auger Observatory, should significantly improve the UHECR statistics in the Northern hemisphere by approaching Auger current statistics at the end of its operating time \cite{TA}.
 This trend, if indeed due to a change in the composition of UHECRs, has extremely important implications. First, it rules out most astrophysical models that predicted a proton dominated composition at the highest energies, and in particular the interpretation of the ankle as being the pair production dip. The proton dominated mixed composition proposed in \cite{Allard05a} which predicts a composition going to lighter elements with increasing energy as well as the classic ankle are also incompatible with this trend. As mentioned in the previous paragraph such an evolution of the composition is not expected if the different species present in the source composition are accelerated above $10^{20}$ eV per nucleon in which case the composition would be expected to become lighter even for heavy dominated compositions at the sources. As discussed in \cite{Denis2008, Denis2009}, the most natural explanation for the opposite trend observed by Auger would be to assume that the UHECR sources (or at least most of them, see below) are not able to accelerate protons well above $10^{19}$ eV but can provide complex nuclei at higher energies with the charge scaling of the maximum energy expected in confinement limited acceleration processes. A mixed composition (but with a low maximum energy for the proton component while keeping the assumption $\rm E_{max}(Z)=Z\times E_{max}(^{1}H)$) appears in principle quite suitable to reproduce this trend. It is obvious, however, that the observed cosmic-ray spectrum cannot be reproduced with, at the same time, the same composition and the same spectral index as in Fig.~\ref{fig:dipmix}b, if one assumes the proton maximum energy to be much lower. In this case a strong inflexion in the spectrum would take place at an energy corresponding to the maximum energy chosen for the dominant proton component. In order to keep the composition identical to the above mentioned mixed composition case one has to consider much harder spectral indices, as show in Fig.~\ref{fig:lowemax}b where a spectral index $\rm \beta=1.6$ and a maximum energy $\rm E_{max}(Z)=Z\times4\,10^{18}$ eV are considered. Alternatively, one can reproduce a similar composition trend while keeping larger spectral indices (see, however, the discussion below) by enriching the composition in heavy elements \cite{Denis2008, Denis2009} as displayed in Fig.~\ref{fig:lowemax}c where the relative abundance of the heavy component is assumed to be 30\% (the main component at the source besides protons), the spectral index $\rm \beta=2.0$, and we kept $\rm E_{max}(Z)=Z\times4\,10^{18}$ eV. In both cases, a good agreement with the observed spectrum is found, and the trend of the composition getting heavier with the energy is reproduced \footnote{These two examples are obviously different from the point of view of the detailed evolution of the composition and might not provide an exact fit of shower depth measurements (they have not been optimized for that purpose) that we consider to be out of the scope of this discussion. Many astrophysical free parameters, that we discuss later,  are not included in these calculations and can be played with to exactly reproduce the data.}. In both cases, the high energy suppression of the flux can be mostly attributed to the GZK effect, as the inflexion in the elemental spectra of complex nuclei are due to their interaction with far-IR photons (that take place at energies lower $\rm E_{max}(Z)$). The case of pure nuclei compositions to explain Auger composition trend has been considered in  \cite{Hooperd2010}. In this case, as discussed above, the maximum energy per nucleon at the sources has to be low to prevent the composition from becoming globally lighter. Also the spectral indices must be hard enough so that for the secondary nucleons are numerous enough to provide a light composition at the ankle while avoiding a strong inflexion of the spectrum at high energies, where the secondary nucleon component disappears. A similar composition trend above the ankle can then be reproduced assuming Si or heavier primaries at the sources. In terms of source models, pure nuclei compositions have of course very different implications from mixed composition models as they would imply the suppression of protons acceleration in UHECR sources at all energies. Let us note that some scenarios have been proposed to accelerate dominantly UHECR heavy nuclei, for instance, in magnetars or GRBs \cite{Epstein, Calvez, Metzger} although there is no strong consensus on the UHECR composition expected to be accelerated by these sources (see for instance \cite{Arons} in the case of magnetars and \cite{Meszaros2008, Kohta2008} for GRBs).
 
On the other hand, a mixed composition model, under the low $\rm E_{max}$ assumption (hereafter low $\rm E_{max}$  models), implies that protons can be significantly present in the source composition (even dominant in the two examples we showed) but that the requirements for their acceleration at the highest energies are not (often) met. In the simple examples displayed in Fig.~\ref{fig:lowemax}b/c, all the sources were assumed to have the same maximum energy above which an exponential cut-off was applied. However, one could argue that a dispersion of the maximum energy at the sources (as proposed by \cite{KS2006} for the dip model) is possible, if not likely. This point is usually illustrated by using the famous Hillas criterion \cite{Hillas}, stating that the Larmor radius of the accelerated particles can not exceed the size of the source ($\rm r_L(E)\leq R_s$). From this argument, one can simply relate the expected maximum energy at the sources with their magnetic luminosity. The simplest estimate \cite{Achterberg2002} gives $\rm E_{max}\sim2.5\times10^{20} \,Z\,\beta_{s}\,\Gamma_{s}\times(L_{B}/10^{46}erg\,s^{-1})^{1/2}$ , where $\rm \beta_s$ and $\Gamma_s$ are the speed and the Lorentz factor of the shock\footnote{Let is note that this is a necessary but not sufficient condition as it is not dealing with energy losses during the acceleration process. Let us note however, adiabatic losses and synchrotron losses would provide larger maximum energy for nuclei than for protons (in the case of synchrotron losses, the scaling of $\rm E_{max}$ with A and Z depends on the energy evolution of the acceleration time). In the case of losses due to interaction at the sources, this discussion is strongly dependent on the source environment \cite{Allard2009}} (see \cite{Lemoine2009} for a more detailed discussion and derivation including a discussion of source beaming and adiabatic losses). From this argument it is often stated that very powerful sources such as FR-II galaxies \cite{Rachen1993} or GRBs \cite{Waxman1995, Vietri1995} are prime candidates for the acceleration of cosmic ray protons above $10^{20}$ eV. These type of sources, as well as very powerful AGNs in general, are,  however, extremely rare in the local universe (e.g, within the horizon of UHECR protons), while the simple luminosity requirement is much looser for nuclei (especially with large Z). It does not thus look unreasonable to assume that sources of protons above $10^{19}$ eV might be outnumbered and dominated in their total contribution to the UHECR flux by less powerful sources that can only reach the highest energies for highly charged heavy nuclei (see discussions in \cite{Ptitsyna2008, Meszaros2009, Hajime2011, Susumu2007}).

It is obviously non-trivial to relate the luminosity of a source (either the bolometric luminosity or the luminosity in a given wavelength range) with its magnetic luminosity and the expected UHECR luminosity. It is, however, interesting to note that typical shapes of the luminosity function of AGNs (either radio loud or radio quiet) have been observed in different wavelengths, going from a relatively flat power law decrease at low luminosity (which implies a positive evolution for the luminosity density) to a much steeper decrease above a given redshift dependent break luminosity (see for instance \cite{Dunlop1990, Boyle2000, Aird2010} and \cite{Longair} for a review and discussion).  The cosmological evolution of their density is also thought to be luminosity dependent with very luminous sources presenting in general much higher evolution with redshift than the faint ones. It is quite tempting, though speculative, to draw a parallel for the case of UHECRs, and to invoke that the decrease of the proton component suggested by Auger composition analyses could be associated with a break of the luminosity function of UHECR sources (as recently discussed in \cite{Carl2011}, affecting nuclei with charge Z at an energy Z times larger), rather than an exponential cut-off for all sources. Let us note that the cosmological evolution of the luminosity function, as well as the evolution of the composition with luminosity and with the redshift, are expected to play a role in the UHECR output. All these unknown astrophysical parameters are actually  somehow embedded into the spectral indices, the maximum energy, or the composition (becoming then "effective" rather than "astrophysical" parameters) in models based on simplifying assumptions that we previously discussed. As a speculation, let us note that this improved version of the low $\rm E_{max}$ could also (to some extend) handle differences in the energy evolution of the air shower depth and its RMS between the two hemispheres, if this were to be confirmed by Telescope Array. Indeed a larger number of sources able to accelerate protons above $10^{19}$ eV in the northern hemisphere could be enough to keep the $\rm X_{max}$ distribution broad (without implying a proton dominated composition as the $\rm X_{max}$ distributions are expected to rapidly broaden with increasing proton fractions) and prevent the clear trend observed in the South being reproduced in the North.

From the point of view of the GCR to EGCR transition, the mixed composition low $\rm E_{max}$ models have the same implications as discussed previously for the mixed composition model. A similar model (e.g, based on a mixed composition and low maximum energy for the proton component) was more recently proposed in \cite{Dis1, Dis2}, the major difference being that in this model the ankle is interpreted as being the transition between a collapsing extragalactic proton component and the extragalactic complex nuclei component. As for the dip model (although the interpretation of the ankle is totally different) the transition from GCR to EGCR has to take place at lower energy and strong composition features (a steep transition from heavy to light and then a break) are in this case expected between $10^{17}$ and $10^{18}$ eV.  

Interestingly, \cite{Lemoine2009} pointed out that an anisotropy pattern at the highest energies could be used to constrain the UHECR source composition. By assuming that a given significant excess of events (with arbitrary angular size) in some region of the sky above a threshold energy E is due to particles of primary charge Z, then the presence or absence of an excess of similar angular extension as or more significant  at an energy E/Z can allow one to estimate or constrain the fraction of protons injected at the source(s)\footnote{The arguably simplistic assumption that an excess of events is due to a single element of charge Z is the price to pay for this statement to be totally independent of the  extragalactic and galactic magnetic fields.}. The Pierre Auger collaboration recently reported the absence of lower energy counterparts to the excess of events around Centaurus A or the excess of correlating events with the AGN of the VCV catalogue reported in \cite{AugerUpdate}, trying different energy thresholds corresponding to different hypotheses for the charge of the primary causing the high energy excess. Many different interpretations of this result are viable, and more statistics are needed to better characterize the anisotropy signal at the highest energies. It is in particular not clear whether some fractions of protons are needed above $5\,10^{19}$ eV to reproduce the observations or, on the contrary, if large metalicities are favored. Let us note as well that, in low $\rm E_{max}$ scenarios, intermediate nuclei like CNO could play a key role in anisotropy patterns. These nuclei are indeed expected to show a much larger attenuation of their diffuse flux than heavier nuclei (due to their smaller horizon at very high energy), but they could, however, reach the Earth relatively unaffected from nearby sources (so with a  larger relative abundance than in the diffuse flux) that are likely to be the main contributors to anisotropy patterns. The study of the energy evolution of the anisotropy patterns is, in any case, a very promising tool for constraining UHECR composition, although these constraints are only valid for the sources that produce the anisotropy patterns. More statistics from the Pierre Auger Observatory and, in the future, from JEM-EUSO are expected to bring more constraints on the origin of UHECRs in the local universe.


\section{Cosmic magnetic fields}

Cosmic magnetic fields are expected to be major actors in the propagation of extragalactic UHECRs. Although our understanding of the Galactic magnetic fields has made some progress in the recent years due to the increasing amount of constraining observations (see for instance \cite{Sun2008}), the extragalactic magnetic field remains rather poorly known outside structures like galaxy clusters. In this section we will restrict our discussion to how the extragalactic magnetic fields could modify the conclusions we reached in the previous section dedicated to the rectilinear regime. We will not discuss the propagation in the Galactic magnetic field but some details on the observations and constraints can be found in \cite{Vallee2005, Han2009, Han2009, Beck2008, Sun2008, Brown2011}, analytical models of the magnetic fields in the Galaxy in \cite{Stanev1997, Harari1999, Prouza2003} and studies on the propagation of UHECR and the expected consequences on their arrival directions in \cite{Stanev1997, Harari1999, Nagataki2003, Prouza2003, Kachelriess2006, Takami2008, Gwen2011}. We will also not discuss in detail the impact of extragalactic magnetic fields on the expected anisotropies of the UHECR sky. Detailed  discussions (including for some of them a discussion on composition or a proposed interpretation of the available data) can be found, for instance, in  \cite{Harari2002,Yoshiguchi2003, Toko2004, Arm2005, Scat2008, Kashti2008, Takami2009a, Aharonian2010, Lemoine2011} and compared with the case of negligible fields studied in \cite{Demarco2006}.

The extragalactic magnetic fields (EGMF) are currently poorly constrained (see \cite{Beck2008} for a review), and are often assumed to be negligible (which is what we implicitly did in the previous section). Indeed, no convincing mechanism that could create strong magnetic fields over very large scales has yet been clearly established. However, the current upper limits do not exclude the presence of magnetic fields of several nG in the extragalactic medium \cite{Kronberg1994, Blasi1999}. In the case of non negligible magnetic fields, several effects on the predicted spectrum can be expected as a result of the existence of magnetic horizons \cite{Stanev2000} evolving with the rigidity of charged particles. Before discussing these effects, it is important to mention that the influence of the extragalactic magnetic fields on the spectrum and the composition of UHECRs can only be discussed in the context of a particular source distribution model (e.g, the source density and luminosity have to be specified). Indeed the influence of a given field configuration does ultimately depends on whether or not the field is strong enough to prevent particles of a given energy and rigidity from reaching the earth within a propagation time shorter than the typical energy loss time or the age of the sources. These considerations obviously depend, for a given field model, on the source distribution and luminosity function, especially in the local universe. In particular,  \cite{Aloisio2004} demonstrated the so-called {\it propagation theorem} stating that, if the distance separation between (identical) sources is smaller than all the typical propagation length scales (e.g,  loss scale, diffusion scale...), the propagated spectrum is independent of the propagation mode. A given magnetic field can then only affect the UHECR spectrum and composition if the source distribution does not fulfill the requirements of the {\it propagation theorem}  at some energies/rigidities.

\begin{figure}[t]
\includegraphics[width=0.5\columnwidth]{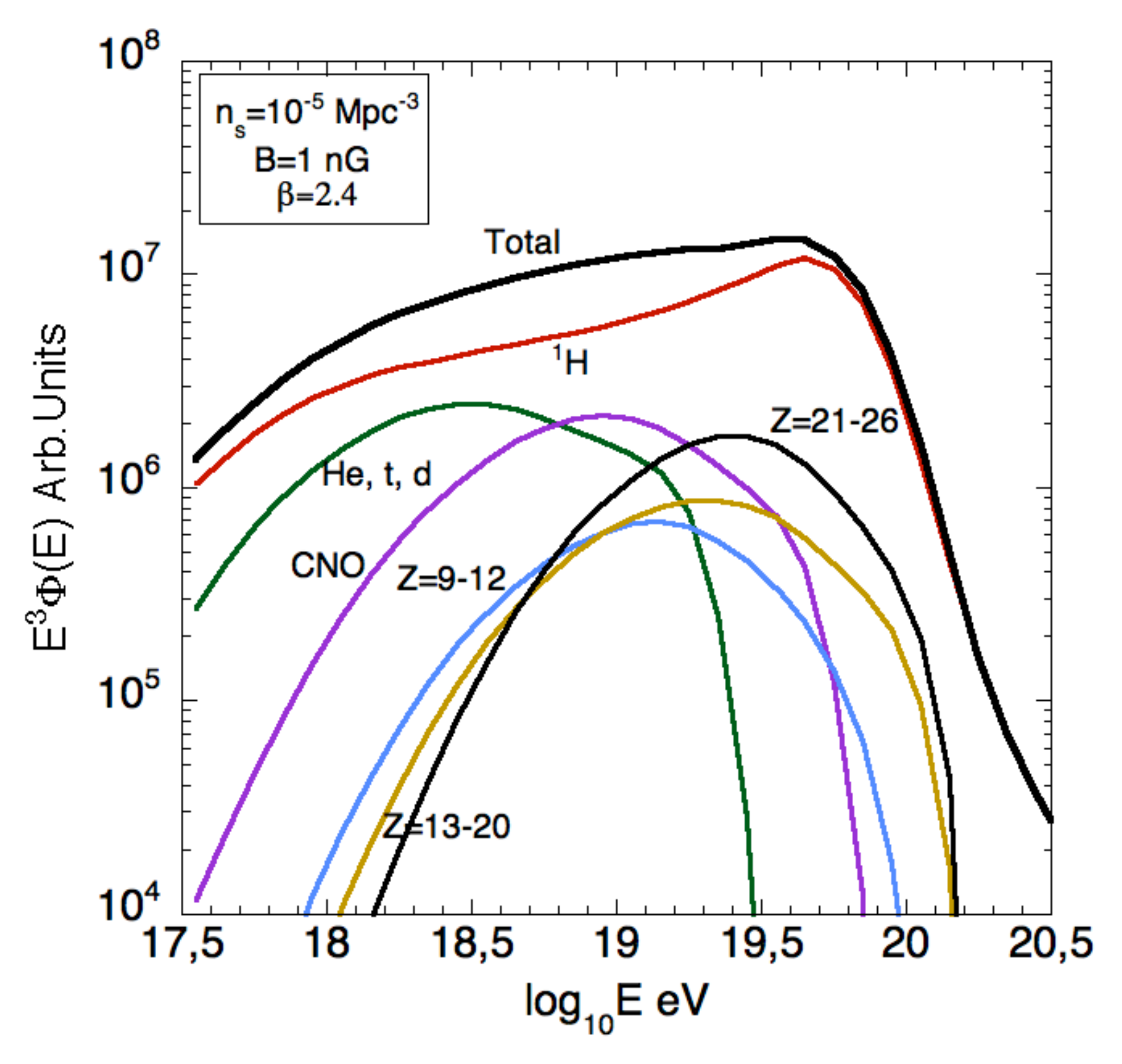}
\hfill
\includegraphics[width=0.5\columnwidth]{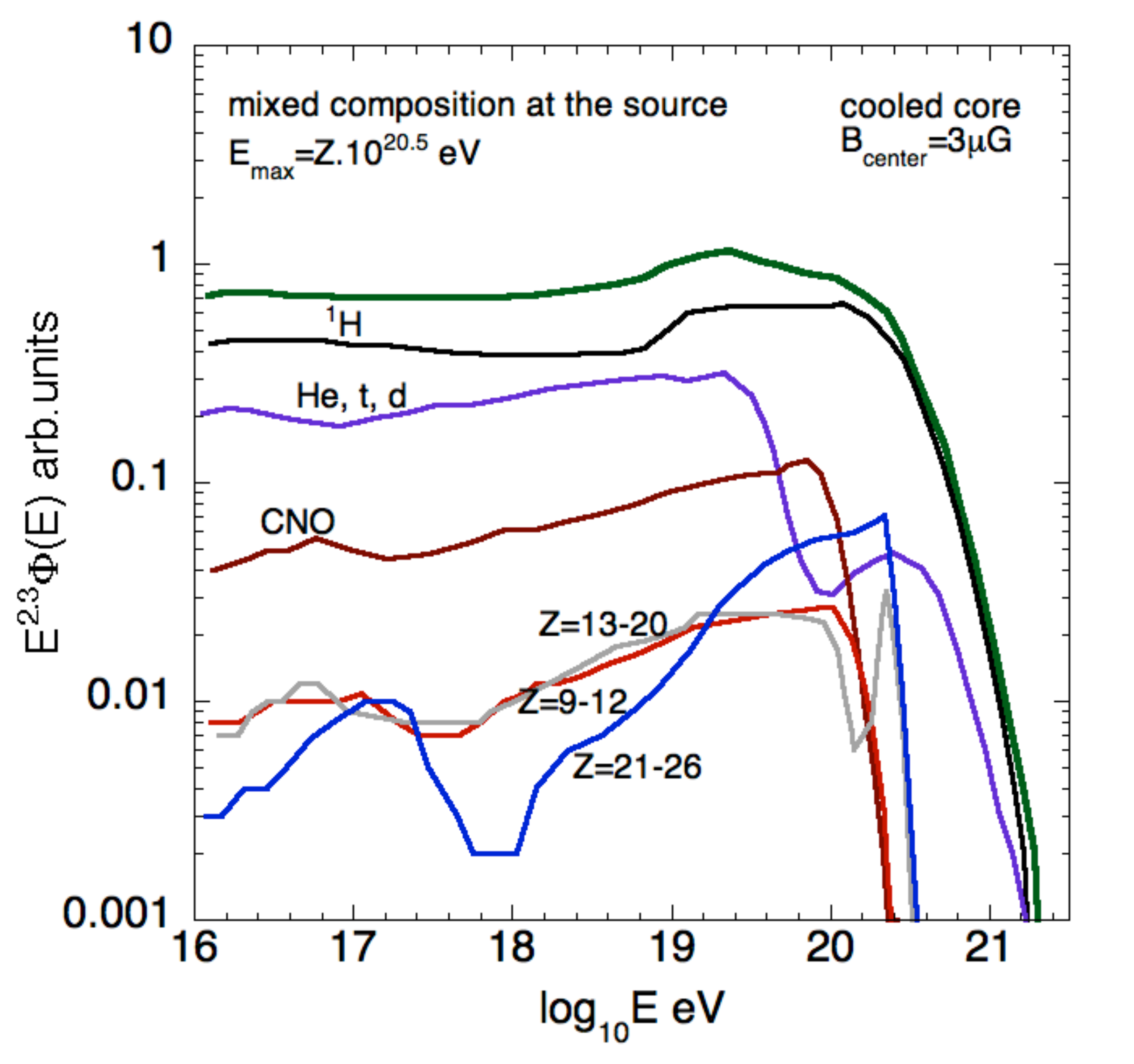}

\caption{Left: Contribution of the different elemental groups to the extragalactic UHECR spectrum assuming a mixed composition in the presence of a homogenous extragalactic magnetic field of 1 nG (Kolmogorov turbulence, $\rm \lambda_{max}=1$ Mpc). Right: Cosmic-ray energy spectrum after  escape from a galaxy cluster environment (e.g, cosmic-rays reaching a distance D= 5 Mpc from the cluster center). The UHECR source is assumed to be at the center of a cool core cluster with a central magnetic field, Bc = 3 $\rm\mu G$. A mixed composition similar to Fig.~\ref{fig:dipmix}b with maximum energy $\rm E_{max}(Z)=Z\times10^{20.5}$ eV is assumed, the contribution of different elemental groups is shown (see labels) .
}
\label{fig:EGMF}
\end{figure}

Strong turbulent magnetic fields, for instance, have been invoked to explain the absence of the GZK cut-off claimed by the AGASA collaboration a few years ago  (see for instance \cite{Olinto1999, Aloisio2004, Del2004}). These type of scenarios, however, are now somewhat less attractive with the revision of AGASA reconstruction methods and the observation of a suppression of the UHECR flux above 3-5$\times 10^{19}$ eV by the HiRes and Auger experiments. Another effect related to the existence of magnetic horizons\footnote{It has actually the same cause as the previous one but requires less extreme configurations from the point of view of the magnetic fields and the source distribution.} may arise if the observer is outside the confinement sphere around a source, which becomes smaller as the energy decreases. In this case, particles emitted by that source, at some rigidity, simply could not reach us\footnote{More correctly, magnetic horizon effects are expected to take place as soon as the lower part of the time of flight distribution of particles at a given energy/rigidity is truncated by the effect of the magnetic field, the highest part of the distribution being controlled by the effect of energy losses or the age of the source}. The induced suppression of the flux at low energy could be of crucial importance (between $10^{17}$ eV and $10^{18}$ eV, or even above in the case of strong fields) and impact the transition between Galactic and extragalactic CRs. Such a scenario has been invoked in the context of a second-knee transition model (e.g, the dip model) to lower the contribution of the extragalactic protons around $10^{17}$ eV. It has been shown \cite{Lemoine2005, Aloisio2005, BereDiff2007} that, for a typical distance between sources of the order of 50 Mpc, a 1 nG field (with a largest turbulence scale $\rm \lambda_{max}$ = 1 Mpc) could significantly suppress the extragalactic proton component below $10^{18}$ eV and thus avoid a too large proton fraction at $10^{17}$ eV. While the above mentioned studies used an homogenous turbulent field,  \cite{Kotera2007} considered more realistic inhomogenous magnetic field configurations, with intensities in the voids between 0.1 and 10 nG, and found only a weak dependence of the  extragalactic proton spectrum above $10^{17}$ eV on the spatial distribution of the field (as controlled by the specific model relating the magnetic field to the matter density). 

For a mixed composition, the departure from the propagation theorem requirements at a given energy obviously depends on the charge of the particles, these requirements becoming more stringent as the charge Z increases. The propagation of a mixed composition in extragalactic turbulent magnetic fields was considered in \cite{Noemie2008}. An illustration of the charge dependance of the magnetic horizon effects is given in Fig~.\ref{fig:EGMF}a, showing the propagated spectrum for a mixed composition (see above) in a 1 nG purely turbulent homogenous extragalactic magnetic field for a realization of a source distribution with density $10^{-5}$ $\rm Mpc^{-3}$. One can see the different elemental groups entering the magnetic horizon at energies increasing with the charge. In particular the effect of magnetic horizons on the proton component is small and restricted to energies below $10^{18}$ eV while the heavy component is almost absent at low energy and totally recovers above $10^{19}$ eV just before being affected by the pair production and photodisintegration losses. The final cut-off of the heavy component takes place earlier than in Fig.~\ref{fig:dipmix}c, mostly due to the low density used in this illustration (there is no source closer than $\sim$ 35 Mpc) and, to a lesser extent, the increase of the time of flight due to the (quite large) magnetic field assumed.  The overall effects of the EGMF on the propagated spectrum are relatively small for a proton dominated mixed composition with a large value of $\rm E_{max}$. The extragalactic composition is lighter at low energy and becomes progressively heavier and closer to the rectilinear case at high energy. The spectrum obtained is overall harder (due to the effect of the magnetic horizons and the suppression of low rigidity particles at low energy) which means that in general softer spectral are required to fit the data. This is all the more true if the relative abundance of metals at the sources is increased compared with the mixed composition model discussed earlier. Let us note that strong magnetic fields could be invoked to strongly suppress a heavy component with respect to protons (even though this scenario does not seem to be supported by data). Indeed, one can imagine source distributions and EGMF configurations for which the minimum time of flight from the sources to the Earth for heavy nuclei is  at all energies larger or comparable to the energy loss horizon\footnote{By this, we mean the largest possible time of flight for a particle to remain above a given energy and a given mass. Numerically this maximum time of flight is model dependent and not equivalent to the loss length used in the previous sections, see \cite{Noemie2008}.}. In the homogenous turbulence case studied in \cite{Noemie2008}. These kind of  effects are obtained for EGMF variances on the order of 10 nG and source densities lower than  $10^{-5}$ $\rm Mpc^{-3}$. In this case, however, the proton component is itself strongly affected below 10 EeV by magnetic horizon effects. These type of mechanisms have been earlier invoked in \cite{Sigl2005, Sigl2007}  in the case of inhomogenous EGMF  to show that the transition scenarios implied in the dip or the mixed composition models could be strongly modified if UHECR sources are immersed in strongly magnetized structures (see below). The magnetic horizon effect we discussed here, mostly qualitatively, are obviously extremely difficult to predict quantitatively as they depend on the very poorly known EGMF configuration and the source distribution and luminosity function. They are in any case most likely to affect the heavy component of the extragalactic UHECRs. 

In the pioneering work on the propagation of UHECRs in the EGMF, a lot of effort has been devoted to studying  the effect of potentially strong fields in the local supercluster (see for instance \cite{Giler1980, Bere1990, Medi1997, Olinto1999, Lemoine1999, Isola2002, Stanev2003}) focusing mostly on the expected anisotropies or the modification of the GZK argument.  More recently, the discussion of the effect of the EGMF enhancement in large scale structures has been undertaken using large structure formation simulations including MHD treatment for the evolution of the magnetic fields \cite{Dolag2002, Dolag2004, SiglMHD2004, Kang2007, Ryu2008, Donnert2009, Ryu2010} (some of these simulations being constrained by the local density velocity field to provide more realistic field configurations in the local universe). These simulations rely on different assumptions on the  origin of the fields and the mechanisms involved in their growth. They are ultimately normalized to the values observed at the present epoch in the central regions of galaxy clusters (see discussion in \cite{Kotera2007}). The outcome of the different simulations strongly differ. In particular, the volume filling factors for strong fields (say above 1 nG) vary by several orders of magnitude from one simulation to the other leading to opposite conclusions on the feasibility of UHECR astronomy, even in the case of proton dominated compositions at the highest energies. While simulations from \cite{Dolag2004} and \cite{Donnert2009} predict very weak fields outside large structures and as a result very low deflections for UHECR protons above a few $10^{19}$ eV, \cite{SiglMHD2004} predicts a much larger influence of strong magnetic fields and deflections of protons around $10^{20}$ eV as large as 20 degrees. Simulations by \cite{Ryu2008} represent an intermediate case with moderate but significant deflections of UHECR protons. In particular, the authors pointed out the possible ambiguities for the source identification from the UHECR arrival direction (this point was made earlier in \cite{Scat2008} with a different method based on magnetic scattering centers in an otherwise non magnetized universe). 

One of the common features of these simulations is the presence of strong fields (of the order of 1$\rm \mu G$ or higher) in galaxy clusters and, in general, stronger fields within matter overdensities as expected from simple considerations. The eventuality that UHECR sources are harbored in these overdense regions is quite likely, at least for some fraction of them. The UHECR spectrum and composition could then be greatly influenced by these magnetized regions, UHECRs would have to go through before experiencing much lower magnetic fields in the intergalactic voids. As mentioned in the previous paragraph, the influence of these magnetized regions possibly surrounding the sources has been first pointed out in \cite{Sigl2005}. With the very flat magnetic field radial profile in the intra-cluster medium  predicted by the cosmological simulations of  \cite{SiglMHD2004}, the authors found that particles with rigidity below $\sim10^{19}$ V were strongly confined and as a result very strong mangnetic horizon effects were expected in the propagated spectrum (depleted below $10^{19}$ eV) and the composition (the intermediate and heavy nuclei being strongly suppressed at all energies). More recently, the escape of UHECRs from galaxy cluster environments was considered in \cite{Kotera2009}. This study was based on the output of the structure formation simulations by \cite{Dubois2008} and considered, besides the effect of the intra-cluster magnetic fields (whose radial attenuation was much steeper than in \cite{SiglMHD2004}), the enhancement of the photon background due to the contribution of galaxies inside the cluster and the contribution of hadronic interactions. The sharper magnetic profile and the addition of the contribution of the cluster environment makes the outcome of the simulation quite different from \cite{Sigl2005, Sigl2007}. The spectrum of UHECRs escaping the cluster environment, assuming the same mixed composition and maximum energy as in Fig.~\ref{fig:dipmix}c, is displayed in Fig.~\ref{fig:EGMF}b. While protons manage to escape at all energies, complex nuclei see their spectrum increasingly modified (hardened) as their mass and charge increase, due to their lower rigidity and larger interaction cross sections at a given energy (see \cite{Kotera2009} for more details). While the proton spectrum is softened at low energies by the addition of secondary nucleons, heavy nuclei are strongly depleted \footnote{The cut-off of the complex nuclei spectra at the highest energies and the secondary nucleons bump above $\sim 5\,10^{18}$ eV is due to interactions with CMB photons and would happen similarly in the intergalactic medium} before recovering at high energies as the confinement time decreases. This illustrates the fact that the spectrum of UHECRs escaping a source environment can significantly depart from the standard hypotheses we made in the previous section. Qualitatively equivalent effects (based on the larger confinement time and interaction probability of heavy nuclei at a given energy in a given environment) could take place in different astrophysical environments (for instance in jets) and lead to a "non-standard" spectrum and composition shape of the UHECR output.

\section{Secondary cosmogenic messengers}

The inevitable existence of secondary particles, a by-product of the interactions of UHECRs with photon backgrounds, was understood quite early after the prediction of the GZK cut-off and the first detailed studies on UHECR propagation. The unavoidable production of UHE neutrinos, through the decay of charged pions, was first pointed out by Berezinsky and Zatsepin \cite{BereOriginal}. Since these {\it cosmogenic neutrinos} can travel from their production site without undergoing interactions or deflections, they were, soon after this pioneering work, considered as potentially interesting probes of the highest energy phenomena in the universe. Their expected flux on earth was intensively calculated in the past decades for different astrophysical scenarios on the cosmological evolution of the sources, the composition or the maximum energy at the sources (see for instance \cite{Stecker1979, Hill85, Engel2001, Kalashev2002, Seckel2005, Hooper2005, Ave2005, Stanev2005, DenisNeut2006, Anchordoqui2007, Takami2009, BereNeut2009, Ahlers2009, Kotera2010}). 

The UHE cosmogenic neutrinos (above say $10^{17}$ eV) are provided  by photopion interactions of the UHECRs with CMB photons, and the expected flux on Earth mostly depends on a few astrophysical parameters such as the cosmological evolution of the CR luminosity (see \cite{Seckel2005}) or the maximum energy at the sources. Indeed, strong cosmological evolutions provide  large weights to distant sources for which interactions with more energetic and denser photon backgrounds take place. On the other hand, the maximum energy at the sources must be large enough for particles above the pion production threshold to be present at all redshifts. The influence of composition \cite{Hooper2005, Ave2005, DenisNeut2006, Anchordoqui2007, Kotera2010} is slightly more complicated to handle, however, for a given maximum energy per nucleon (of course above the pion production threshold) the neutrino flux predicted for heavy primaries (say iron) is  lower but comparable with the expectations for proton primaries. The difference depends mostly on the spectral index required to fit the UHECR spectrum (the neutrino deficit for heavy primaries is larger when a soft spectral index is needed) and on the required total injected CR luminosity (which is usually lower in the case of heavy primaries, due to the lower energy losses they experience below $10^{19}$ eV, see Fig~.\ref{fig:lossnuclei}b and discussion in \cite{Decerprit2011}). At PeV/multi-PeV energies cosmogenic neutrinos are produced by interaction with the IR/Opt/UV backgrounds \cite{Stanev2005}. The production is  strongest for models requiring soft spectral indices like the dip model (unless a low energy cut mechanism is applied \cite{DenisNeut2006}).

\begin{figure}[t]
\includegraphics[width=0.5\columnwidth]{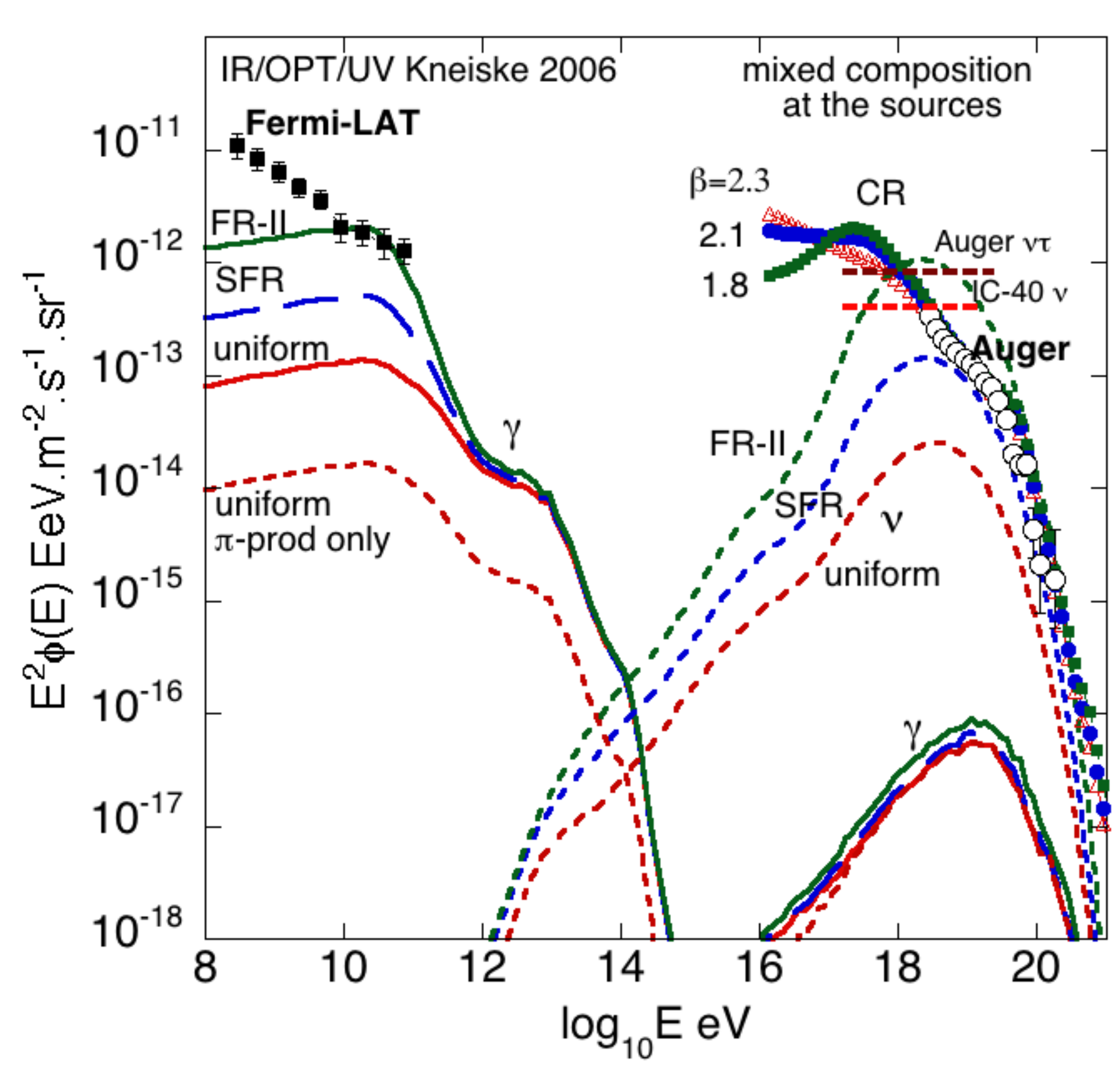}
\hfill
\includegraphics[width=0.5\columnwidth]{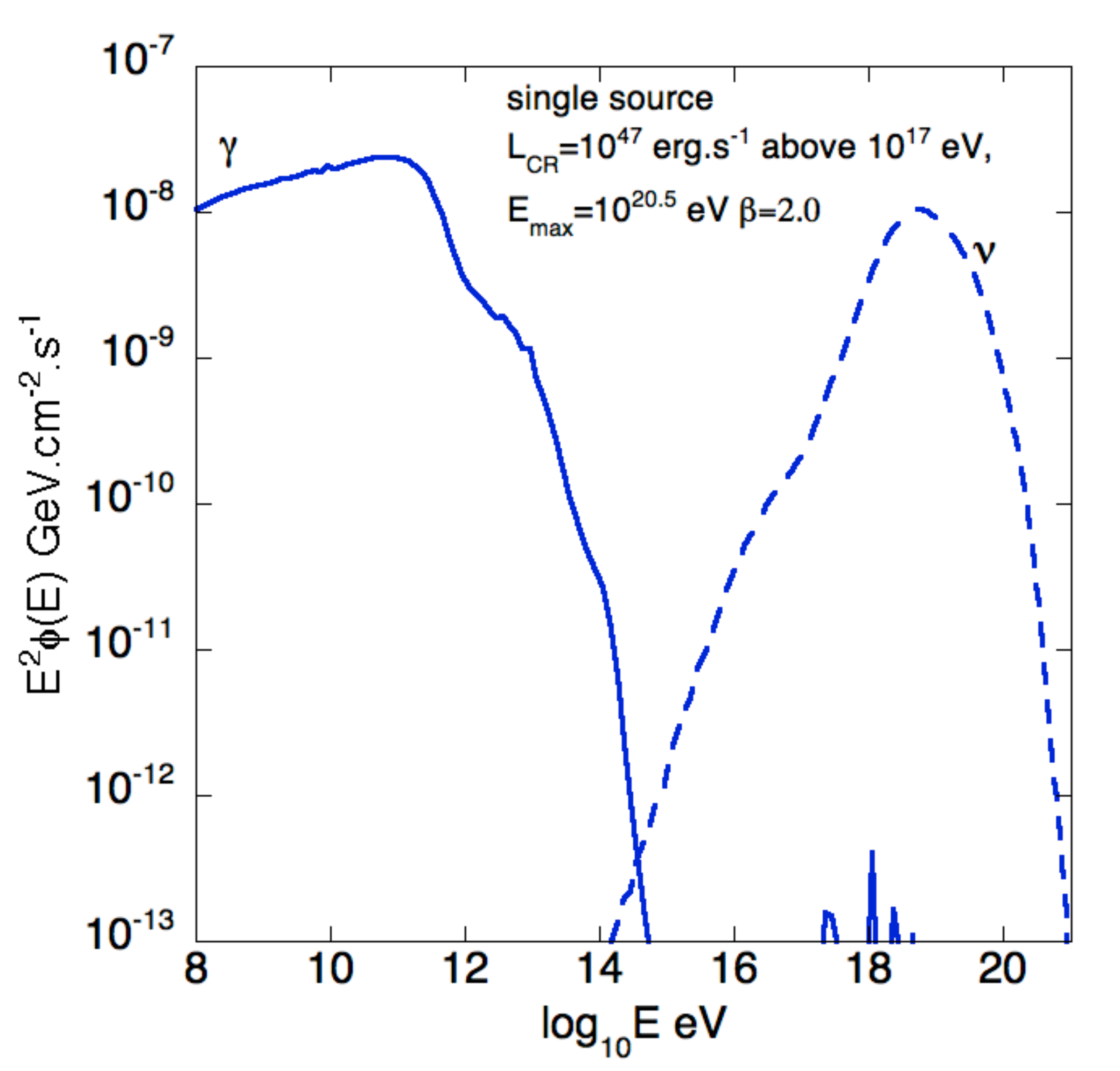}

\caption{Left : Cosmic ray (markers), all neutrino flavors (dashed lines) and photon (solid lines) spectra ($\rm E^2\times dN/dE$) for the mixed composition model ($\rm E_{max}=Z\times10^{20.5}$ eV), compared to the Auger spectrum (\cite{Augersp2010}; open circles) and Fermi diffuse gamma-ray spectrum (\cite{FermiGRBack10}; black squares). The contribution of the pion mechanism to the photon spectrum is also shown (dashed lines). The spectral indexes chosen are $\rm \beta=2.3$ for the uniform case (no evolution), 2.1 for SFR and 1.8 for FR-II. The latest Auger 90\% C.L integrated upper limit for tau neutrinos (\cite{AugerNeut2011}, multiplied by 3 assuming a complete mixing of neutrino flavors) as well as the IceCube40 90\% C.L all flavors integrated upper limit  \cite{IceCube2011} are also shown for comparison. Right: Cosmogenic neutrinos (dashed line) and photons (solid line) from UHECRs emitted by a source located at 1 Gpc with luminosity $\rm L_{CR}=10^{47}$ $\rm erg\,s^{-1}$ above $10^{17}$ eV, $\rm E_{max}=10^{20.5}$ eV, and spectral index $\rm \beta=2.0$.}
\label{fig:multi}
\end{figure}

Cosmogenic gamma-rays are also produced during the propagation of UHECRs. Unlike neutrinos, these very high energy gamma rays interact rapidly and produce electromagnetic cascades. As a result, the universe is opaque to gamma-rays from a few hundreds of GeV to a few $10^{18}$ eV. Above $10^{19}$ eV the universe becomes more and more transparent to photons and very high energy gamma rays can propagate a few tens of megaparsecs without losing a large amount of energy. As a result, these very high energy cosmogenic gamma-rays were discussed in the literature either as signatures of the so called Top-Down models (see for instance  \cite{Protheroe1996, Lee1998, Sigl1999, Semikoz2004}) or as probes of UHECR acceleration in the local universe (e.g, \cite{Yoshida1993, Protheroe1996, Lee1998,  Semikoz2004, Gelmini2007a, Gelmini2007b, Taylor2009, Taylor2010, kuempel2009, Hooper2011, Ahlers2011, Decerprit2011}).

Since electromagnetic cascades, piling up below 100 GeV, are produced during UHECR propagation and are likely to be associated with the production of cosmogenic neutrinos, it has been soon realized that measurements of the diffuse gamma-ray background could allow to put constraints on the cosmological evolution of the UHECR luminosity \cite{Strong73} and on the maximum allowable cosmogenic neutrino fluxes \cite{Bere1975}. Modern versions of these calculations were attempted using the EGRET measurements \cite{EGRET1, EGRET2}  in \cite{Kalashev2002, Semikoz2004, Kalashev2007}. More recently, the Fermi satellite measurements  \cite{FermiGRBack10} reported a gamma-ray background between 100 MeV and 100 GeV lower than previously estimated using EGRET data. The additional constraints brought by this new measurement were discussed in \cite{BereFermi2010, Ahlers2010} and more recently in \cite{Wang2011, Decerprit2011}. The Fermi estimate of the diffuse gamma-ray background brings constraints on the most optimistic scenarios in terms of UHE neutrino fluxes involving UHECR produced by strongly evolving source (for instance the FR-II we mentioned earlier). For lower but significant evolution scenarios, the constraints implied by the diffuse gamma-ray flux depend on the energy transferred to electromagnetic cascades via the pair production mechanism. The softer spectral indices required for the dip model imply a larger energy transfer to $\rm e^+/e^-$ pairs and, ultimately, a larger contribution to the diffuse gamma-ray background. As a result the diffuse gamma-ray flux predicted for the dip model in the SFR evolution case practically hugs Fermi measurements\footnote{Unless one invokes a low energy cut mechanism.}. For other astrophysical models (like the mixed composition or classic ankle models) the constraints are limited to very strong evolutions that are now also constrained by Auger \cite{AugerNeut2011} and IceCube \cite{IceCube2011} upper limits.

This situation is summarized in Fig.~\ref{fig:multi}a, where the cosmogenic neutrino and gamma-ray fluxes for the mixed composition model displayed in Fig.~\ref{fig:dipmix} are shown for the same three source evolution scenarios (and assuming $\rm E_{max}(Z)=Z\times10^{20.5}$ eV). One can see the wide range of UHE neutrino fluxes implied by the different cosmological evolution scenarios. The largest evolution models are constrained (for our assumption on $\rm E_{max}$) by their diffuse multi-GeV gamma-ray fluxes and, more recently, by their UHE neutrino fluxes. In the next 3 years the complete IceCube detector should be able to provide upper limits down to the level of the predictions for the SFR evolution scenario \cite{IceCube2011},  whereas the uniform case may be difficult to constrain even after many years. The non-observation of UHE neutrinos would then not necessarily imply, by itself, that protons are not dominant at the highest energies \cite{BereNeut2009}, but would strongly constrain their cosmological evolution. The UHE neutrino fluxes we presented on this figure are pretty generic fluxes when the maximum energy at the sources is above the pion production threshold. It would not be the case anymore for low $\rm E_{max}$ models, at least in their simplest version, as the absence of sources accelerating cosmic-rays above the pion production threshold with CMB photons would result in a suppression of the UHE neutrino flux whatever the source evolution scenario assumed. However, if one assumes a scenario where strongly evolving UHE proton sources exist but are rare and outnumbered by less efficient accelerators in the local universe (as discussed above), then detectable diffuse neutrino fluxes can still be expected. For instance, a contribution to the UHECR spectrum as low as $\sim$10\% at $10^{19}$ eV from sources accelerating protons above $10^{20}$ eV with a cosmological evolution similar to the FR-II scenario would be enough to produce almost as many UHE neutrinos as the SFR case shown in Fig.~\ref{fig:multi}. This is obviously one of the most optimistic cases one could think of for low $\rm E_{max}$ models. In the any case, if the composition trend suggested by the current Auger composition analyses is confirmed, then the observation of  such a diffuse neutrino flux would certainly represent a signature of the existence of remote UHE proton accelerators.

For scenarios where the acceleration of protons would only be possible in very rare and powerful sources (that could  very well be mostly outside the UHECR horizon) the potential detection of point-like (or slightly extended) cosmogenic neutrinos or GeV-TeV gamma-ray \cite{Ferrigno2005, Arm2006} fluxes could also be a promising way to reveal the presence of these powerful accelerators. In the case of gamma-rays, two types of signatures can be expected depending on the magnetic fields in the vicinity of the source or in the extragalactic medium. The first category of gamma-ray signals could be due to pion decay and pair production induced electromagnetic cascades, which, however require very low extragalactic magnetic fields (well below $10^{-12}$ G \cite{Gabici2005}) to remain within small angular size and easily detectable by gamma-ray observatories. The second type of gamma-ray signals could be provided by synchrotron radiation, from the pion decay and pair production induced UHE secondary pairs, in the magnetized environment of a source \cite{Gabici2005, Gabici2007, Gabici2011}. In these cases, the detection is most promising for extremely powerful sources located around 1 Gpc, which are less constrained by their contribution to the UHECR flux  and would keep an angular size well below 1 degree \cite{Kotera2011}. Interestingly, GeV-TeV cascades were recently proposed  in \cite{Essey10} (and recently re-examined in \cite{Ahlers2011}) as a possible interpretation of  the TeV signal observed by HESS in the direction of the AGN 1ES0229+200 (z=0.14). Assuming a point source cascade signal, Essey et al. estimated the implied source luminosity in cosmic ray protons above $10^{16}$ eV to be between $\sim10^{46}$ and $10^{49}$ $\rm erg\,s^{-1}$ (depending on the maximum energy assumed). If confirmed, this interpretation would not be an unambiguous signature of the acceleration of particles above the pion production threshold but would imply either extremely weak magnetic fields ($\leq10^{-14}$ G) on very large scales or  even greater source luminosities (to allow the same flux to be within the HESS point spread function).

Very powerful sources could also produce large point-like (or quasi point-like),  fluxes of cosmogenic neutrinos (see recently \cite{Essey10, Decerprit2011}) due to the fact that, above the pion production threshold with CMB photons, most of the interactions take place within say 10-20 Mpc from the source. Neutrino fluxes reaching $\rm10^{-8}GeV\,cm^{-2}\,s^{-1}$ (which might be detectable during IceCube operating time) at $\sim10^{18}$ eV could be produced by a source of luminosity $\rm L_{CR}=10^{47} erg\,s^{-1}$ located at 1 Gpc (see Fig.~\ref{fig:multi}b). Distant and extremely powerful sources might be prime targets for the current generation of neutrino telescopes as neutrino fluxes of this magnitude from closer sources would almost necessarily imply (unless very strong effects of the extragalactic magnetic fields) UHE cosmic-ray fluxes in excess of what is observed (see discussion in \cite{Decerprit2011}).

\section{Conclusion}

Despite a large increase of the statistics at the highest energies in the recent years, the origin of UHECRs is still far from being understood. This is certainly due to the fact that the richness and complexity of UHECR astrophysics, potentially involving many of the different astrophysical parameters we tried to discuss (at least some of them) in the previous sections, are much greater than what was anticipated in the past decades when data was too sparse. 

Progresses of our understanding can, however, be reasonably expected in the next few years with the increase of the statistics of high resolution UHECR measurements provided by the Pierre Auger Observatory and Telescope Array. While Telescope Array should (to some extent) settle the question of the composition above $10^{19}$ eV in the northern hemisphere, the Pierre Auger Observatory will also accumulate more statistics above $5\,10^{19}$ eV where some hints of an anisotropic UHECR sky have already been pointed out. Detailed studies of the energy evolution of these anisotropy signals will be of crucial importance. Let us note that both observatories are developing "low energy extensions"  \cite{Augerle2011, TA} which should be of great help in solving the longstanding problem of the GCR to EGCR transition.  

The emergence of new experiments such as JEM-EUSO should also be a great step forward by providing detailed sky maps at energies where the statistics are currently extremely low, even above $10^{20}$ eV. With an integrated exposure potentially approaching $\rm 10^{6}$ $\rm km^2\,sr\,yr$ at the end of the operating time (which is $\sim$ 30 times the current integrated exposure accumulated by UHECR experiments), JEM-EUSO should provide an unprecedented view of the UHECR sky (with a full sky coverage) and in particular put constraints above $3\,10^{20}$ eV where UHECR nuclei are expected to vanish from UHECR composition for reasonable assumptions on the distance of the extragalactic sources.

Gamma-ray and neutrino observatories are now reaching sensitivities high enough to constrain some aspects of the origin of UHECRs. The unprecedented sensitivity of CTA \cite{CTA} could allow to detect the first gamma-ray signature of the most powerful UHECR accelerators, either by the GeV-TeV signal provided by electromagnetic cascades initiated by cosmogenic pairs or photons or the synchrotron signal emitted in the magnetized environment of the sources. The IceCube experiment should be able to challenge reasonable expectations for the diffuse cosmogenic neutrino flux in the next few years and bring constraints on the contribution of strongly evolving bright sources to the (acceleration of) UHECR protons above $10^{20}$ eV, even in the context of the low $\rm E_{max}$ models discussed in the previous sections. Quasi point like cosmogenic neutrino fluxes could also be detected in the case of distant and extremely powerful sources. The European project KM3Net \cite{km3}, currently under development, might provide a similar or even greater sensitivity and complete the sky coverage of neutrino observatories. These multi-messenger observations will be crucial to constraint the acceleration of UHECRs (especially proton accelerators above $10^{20}$ eV which are possibly rare in the local universe) beyond the GZK horizon and at remote cosmic epochs.  

As a conclusion, even though the current observations at the Pierre Auger Observatory (e.g, the evolution of the composition and the low level of the anisotropy signal) might question the feasibility of a trivial "point and shoot" UHECR astronomy (at least below $10^{20}$ eV), the possibilities and perspectives of UHECR astronomy using all the different components of multi-messenger high energy astrophysics (each of them bringing a different but crucial piece of the puzzle) remain extremely attractive.


\end{document}